\documentclass[11pt]{article}
\usepackage{graphicx}
\usepackage{amsmath}
\usepackage{amssymb}
\usepackage{tabularx} 
\usepackage{caption}
\usepackage{subcaption}

\begin{document}

\bibliographystyle{OurBibTeX}


\begin{center}
{ \Large  \bf Constraining Grand Unification \\ using first and second generation sfermions }
\\[10mm]

D.~J.~ Miller$^a$ 
\footnote{e-mail: \texttt{David.J.Miller@Glasgow.ac.uk}} 
A.~P.~Morais$^a$
\footnote{e-mail: \texttt{a.morais@physics.gla.ac.uk}}
P.~N.~Pandita$^b$ 
\footnote{e-mail: \texttt{pandita@iucaa.ernet.in}} 
\\[3mm]
{\small \it
$^a$ SUPA, School of Physics and Astronomy, University of Glasgow, Glasgow, G12 8QQ, UK \\[1mm]
$^b$ Department of Physics, North  Eastern  Hill University,  Shillong 793022, India }\\[1mm]
\end{center}
\vspace*{0.75cm}

\begin{abstract}
\noindent
We investigate the spectrum of supersymmetric grand unification models based on the gauge groups $SU(5)$, $SO(10)$ and $E_6$, paying particular attention to the first and second generation. We demonstrate how the measurement of the first or second generation sfermion spectrum may be used to constrain the underlying grand unification structure. The smallness of first and second generation Yukawa interactions allows us to perform an analytic analysis, deriving expressions for the high scale parameters in terms of the low scale sfermion masses.  We also describe a sum rule that provides an $SO(10)$ mass prediction, distinct from $SU(5)$, and discuss $E_6$ models, both with and without extra exotic matter at low energies. The derived relations are compared with numerical results including two-loop running and the full Yukawa dependence. 
\end{abstract}


\setcounter{footnote}{0}

\section{Introduction}

Recent searches for squarks and gluinos at the Large Hadron Collider (LHC) \cite{ATLAS_susy,CMS_susy} have greatly constrained the parameter space of low energy supersymmetry. These searches have mainly been done within the context of the constrained Minimal Supersymmetric Standard Model (cMSSM, for a review see \cite{Martin:1997ns}) where the soft supersymmetry breaking masses are unified at the energy scale where a Grand Unified Theory (GUT) is presumed to exist. Within this framework, the experiments find that squarks must be heavier than about $1.5\,$TeV and gluinos above about 850 GeV to remain unobserved at the LHC.

While these results certainly put pressure on the cMSSM, there is still plenty of room for the discovery of supersymmetry at the LHC, particularly if one is willing to allow supersymmetry to have a relatively heavy spectrum. The desire to keep the supersymmetric spectrum light is driven by the desire for supersymmetry to be the solution of the hierarchy problem, using top squark loops to cancel the quadratic divergence of the Higgs mass arising from top quark loops. The remaining uncancelled logarithmic divergence will again require fine tuning if the stops become too heavy. Of course, this left over {\it little hierarchy problem} is still vastly less problematic than the required fine tuning of the Standard Model (SM) Higgs sector. Furthermore, searches for the third generation squarks remain relatively weak \cite{ATLAS_3g}. One could imagine a supersymmetric model where the first two generations are relatively heavy, avoiding the current LHC constraints, but the third generation is still rather light, diluting the required fine tuning. Indeed, such a scenario is perfectly reasonable even for GUT constrained supersymmetry, which has no {\it a priori} requirement for a common supersymmetry breaking mass scale across the generations. While vastly differing scales would be difficult to generate using the same mechanism, hierarchies of a few orders of magnitude should not be surprising (and are present already in the SM masses).

Irrespective of the details of the mass spectrum, it is still not unreasonable to suppose that our first sight of supersymmetry will be the discovery of squarks and gluinos with masses of a few TeV. If a hierarchy between the generations does exist, the exclusion limits set by the LHC would be weakened~\cite{Baer:2012uy} since only one generation of squarks could be available to produce instead of two. If this generation were the second generation (that is, an inverted hierarchy with the undetected third generation the lightest and the first generation the heaviest) the limits would be further reduced since one could not rely on the valence content of the proton to enhance squark production. After such a discovery, our task will be to examine the supersymmetric spectrum in detail, determine the underlying mechanism for supersymmetry breaking (see for example~\cite{Faraggi:1991bb}) and hopefully build a new theory that explains some of the unanswered questions of the SM.

One such question is, why is the SM built upon the gauge structure $SU(3) \times SU(2) \times U(1)$? Is this a remnant of some larger simple group that is spontaneously broken at a high scale~\cite{Georgi:1974sy}? A GUT scenario of this type is strongly motivated by the running gauge couplings, which within supersymmetric models appear to have a common value at a scale of about \linebreak[4] $M_{GUT} = 2 \times 10^{16}\,$GeV~\cite{Dimopoulos:1981yj}. Fixing two of the gauge couplings by experiment, the third becomes a successful prediction (or rather postdiction) of supersymmetric unification. 

The most popular candidates of a unified gauge group, are the the rank four, five and six groups, $SU(5)$, $SO(10)$ and $E_6$ respectively (for a review see \cite{Ramond:1979py}), and one expects this underlying gauge structure should leave an imprint on the low scale mass spectrum. In this paper, we will examine how we may determine, or constrain, this choice of the underlying group using only the first or second generation of squarks and sleptons, and the accompanying gauge sector. To do this, we will use the renormalization group equations (RGEs) for the first and second generations, which allows us to neglect Yukawa couplings and analytically integrate the one-loop RGEs. When we have  a unification group of rank higher than four (the rank of the SM gauge group), the breaking mechanism generates extra D-term contributions to the soft SUSY breaking scalar masses \cite{Kolda:1995iw}. This is the case of $SO(10)$ and $E_6$ and these D-terms will be included in our analysis. 

For $SU(5)$, the fermions (and accompanying sfermions) are embedded in a $\mathbf{10} \oplus \mathbf{\overline{5}}$ representation, so their soft scalar (GUT scale) masses take on values of $m_{\mathbf{10}}$ or $m_{\mathbf{\overline{5}}}$, depending on which representation they occupy. This is in contrast to the cMSSM~\cite{Kane:1993td}, where all sfermions share a common universal scalar mass $m_0$. For $SO(10)$, although we have a single $16$ dimensional irreducible representation (irrep) for all the sfermions, $\mathbf{16}$, with a common scalar mass $m_{\mathbf{16}}$, the group is rank-5 and must be broken to the rank-4 gauge group of the SM. The subsequent rank reduction provides D-term contributions splitting the soft masses at the GUT scale. $E_6$ is particularly interesting since it is the largest gauge group that supports the chiral structure of weak interaction in four dimensions. In this model all the matter is contained in a single irrep, $\mathbf{27}$, but the large representation provides left over room for extra states. These states could be left at the high scale, and therefore not contribute to the Electroweak (EW) scale spectrum, or may survive down to low scales and be potentially discoverable at the LHC. We will consider both cases here. We will explore all these different possibilities for the underlying theory being broken at the GUT scale and its consequences for the observable sfermions masses at the EW scale.

Generally, in GUT scenarios, extra super-heavy gauge bosons also arise from the adjoint representation of the unified gauge group, and may mediate baryon number violating interactions causing proton decay. However, the supersymmetric GUT scale is considerably higher than that of a non-supersymmetric scenario, causing proton decay through gauge boson interactions to become sufficiently suppressed. Proton decay may still be problematic however, due to the presence of higher dimensional operators \cite{proton_decay}. In this paper, we will assume that this problem is solved by some unknown mechanism at the GUT scale, such as embedding the model in higher dimensions \cite{higher_dim}. Furthermore, we will assume that contributions from colored triplets arising from higher dimensional Higgs representations are absent due to similar considerations that solve the doublet-triplet splitting problem \cite{Sakai:1981gr}.

This paper is organized as follows. In Section \ref{sec:rge}, the analytic solution of the first and second generation sfermion mass RGEs is described, allowing us to define the various coefficients and parameters. In Section \ref{sec:boundary_conditions}, we apply different boundary conditions to the sfermion masses, and show how the determination of the soft masses can be used to distinguish between different supersymmetric grand unification scenarios. The $SU(5)$, $SO(10)$ and $E_6$ boundary conditions are considered, and we will also describe how the different GUT boundary conditions provide further constraints. Since the $E_6$ $\mathbf{27}$-plet has room for additional exotic matter, in Section \ref{sec:e6ssm} we will also discuss an example of an $E_6$ inspired model containing additional matter at the low scale, the E$_6$SSM \cite{e6ssm}. In Section \ref{sec:softsusy} we demonstrate that the sum rules are robust to the inclusion of Yukawa couplings and two-loop effects by comparing with the program SOFTSUSY~\cite{Allanach:2001kg}. In section~\ref{sec:higgs} we will briefly discuss the fine-tuning problem in the light of the recent observation of a Higgs candidate at the LHC, and conclude in Section \ref{sec:conc}.

\section{Integration of the Renormalization Group Equations}
\label{sec:rge}

In this section we will reproduce the analytic scale dependence of the first or second generation scalar masses, as well as some sum rules that are applicable independently of the choice of high scale boundary conditions. These results are largely available in the literature (see for example Ref.~ \cite{Martin:1993ft}) so we include them here for completeness and to set the notation for the discussion to come. 

Neglecting Yukawa couplings, the RGEs for the first or second generation scalar masses in the MSSM \cite{Martin:1993ft,rges}, to one-loop accuracy, are
\begin{eqnarray}
16 \pi^2 \frac{dm^2_{\tilde{Q}_L}}{dt} &=& -\frac{32}{3} g^2_3 M^2_3 - 6 g^2_2 M^2_2 - \frac{2}{15} g^2_1 M^2_1 + \frac{1}{5} g^2_1 S,\label{eq:rg1}\\
16 \pi^2 \frac{dm^2_{\tilde{u}_R}}{dt} &=& -\frac{32}{3} g^2_3 M^2_3 - \frac{32}{15} g^2_1 M^2_1 - \frac{4}{5} g^2_1 S,\label{eq:rg2}\\
16 \pi^2 \frac{dm^2_{\tilde{d}_R}}{dt} &=& -\frac{32}{3} g^2_3 M^2_3 - \frac{8}{15} g^2_1 M^2_1 + \frac{2}{5} g^2_1 S,\label{eq:rg3}\\
16 \pi^2 \frac{dm^2_{\tilde{L}_L}}{dt} &=& - 6 g^2_2 M^2_2 - \frac{6}{5} g^2_1 M^2_1 - \frac{3}{5} g^2_1 S,\label{eq:rg4}\\
16 \pi^2 \frac{dm^2_{\tilde{e}_R}}{dt} &=& - \frac{24}{5} g^2_1 M^2_1 + \frac{6}{5} g^2_1 S,
\label{eq:rg5}
\end{eqnarray}
where $t \equiv \log(Q/Q_0)$, with $Q$ the energy scale of interest and $Q_0$ the unification scale, for which we will use $Q_0=1.9 \times 10^{16}\,$GeV throughout. $M_{1,2,3}$ are the gaugino masses corresponding to the usual $g_{1,2,3}$ gauge couplings, with RGEs
\begin{equation}
8 \pi^2 \frac{dM_i}{dt} =  b_i g_i^2 M_i, \qquad \qquad b_i = \left( \frac{33}{5},1,-3 \right).
\label{eq:rgMi}
\end{equation}
These are identical to the RGEs appearing in Ref.~\cite{Martin:1993ft}, though our convention for $b_i$ differs by a sign.

$S$ is only non-zero if the sfermion masses are not universal at the GUT scale; it is given by~\cite{Ananthanarayan:2004dm, sparticle_GUT}
\begin{equation}
S \equiv Tr(Ym^2) = m^2_{H_u} - m^2_{H_d} +\displaystyle\sum\limits_{\rm generations} \left( m^2_{\tilde{Q}_L} - 2m^2_{\tilde{u}_R} + m^2_{\tilde{d}_R} - m^2_{\tilde{L}_L} + m^2_{\tilde{e}_R} \right).
\label{eq:S}
\end{equation}
Notice that the sum over generations results in $S$ also depending on the third generation soft scalar masses, and therefore implicitly on the third generation Yukawa couplings, which cannot be neglected. However, when constructing the evolution equation for $S$ from the above definition, one finds that these Yukawa couplings cancel (for the same reason that the gravitational anomaly cancels), so an analytic solution is still possible. Indeed, only terms proportional to $S$ itself survive and one finds
\begin{equation}
\frac{dS}{dt} = \frac{66}{5} \frac{\alpha_1}{4 \pi}S \quad \Rightarrow \quad S(t)=S_0\frac{\alpha_1(t)}{\alpha_1(0)}
\label{eq:Sevolution}.
\end{equation}
Here $S_0 \equiv S(0)$ is the value of $S$ at the GUT scale and $\alpha_1 = g_1^2/4\pi$ as usual. 

The absence of Yukawa and trilinear couplings allows equations (\ref{eq:rg1}) to (\ref{eq:rg5}) to be solved analytically. Furthermore, since only gauge interactions contribute to the running, if  the sfermion squared mass-matrices are flavour-blind at the input scale, the squared masses of the gauge-eigenstates for the first two generations will remain diagonal at the supersymmetry breaking scale, with nearly degenerate left/right masses given by
\begin{equation}
\mathcal{L}_{mass} = -
\begin{pmatrix}
\varphi^{*}_L  & \varphi^{*}_R
\end{pmatrix} 
\begin{pmatrix}
m^2_{\varphi_L} + \Delta_{\varphi_L} & 0 \\
0 & m^2_{\varphi_R} + \Delta_{\varphi_R}
\end{pmatrix}
\begin{pmatrix}
 \varphi_L \\
 \varphi_R
\end{pmatrix}.
\label{eq:gauge_space}
\end{equation}
Here $\varphi_{L/R}$ represents any left/right-handed squark or slepton of the first two generations. $\Delta_{\varphi_{L,R}}$ is a D-term contribution arising from the breaking of the Electroweak symmetry, \linebreak $SU(2)_L \otimes U(1)_Y \rightarrow U(1)_{em},$
\begin{equation}
\Delta_{\varphi_{L,R}} = M^2_Z (T_{3\varphi_{L,R}} - Q_{\varphi_{L,R}}\sin^2 \theta_W)\cos 2\beta~,
\label{eq:DeltaEW}
\end{equation}
where $M_Z$ is the $Z$-boson mass, $T_{3\varphi_{L,R}}$ the third component of the weak isospin, $Q_{\varphi_{L,R}}$ the electric charge and $\tan \beta = v_u/v_d$ with $v_u$ and $v_d$ the up-type and down-type Higgs vacuum expectation values (vevs) respectively.
The solution of equations (\ref{eq:rg1}) to (\ref{eq:rg5}) is given by \cite{Martin:1993ft,Ananthanarayan:2004dm}
\begin{eqnarray}
m^2_{\tilde{u}_L}(t) &=& m^2_{\tilde{Q}_L}(0) + C_3 + C_2 + \frac{1}{36} C_1 + \Delta_{u_L} - \frac{1}{5}K,\label{eq:sol1}\\
m^2_{\tilde{d}_L}(t) &=& m^2_{\tilde{Q}_L}(0) + C_3 + C_2 + \frac{1}{36} C_1 + \Delta_{d_L} - \frac{1}{5}K,\label{eq:sol2}\\
m^2_{\tilde{u}_R}(t) &=& m^2_{\tilde{u}_R}(0) + C_3 + \frac{4}{9} C_1 + \Delta_{u_R} + \frac{4}{5}K,\label{eq:sol3}\\
m^2_{\tilde{d}_R}(t) &=& m^2_{\tilde{d}_R}(0) + C_3 + \frac{1}{9} C_1 + \Delta_{d_R} - \frac{2}{5}K,\label{eq:sol4}\\
m^2_{\tilde{e}_L}(t) &=& m^2_{\tilde{L}_L}(0) + C_2 + \frac{1}{4} C_1 + \Delta_{e_L} + \frac{3}{5}K,\label{eq:sol5}\\
m^2_{\tilde{\nu}_L}(t) &=& m^2_{\tilde{L}_L}(0) + C_2 + \frac{1}{4} C_1 + \Delta_{\nu_L} + \frac{3}{5}K,\label{eq:sol6}\\
m^2_{\tilde{e}_R}(t) &=& m^2_{\tilde{e}_R}(0) + C_1 + \Delta_{e_R} - \frac{6}{5}K, \label{eq:sol7}
\end{eqnarray}
where we have
\begin{equation}
   C_i(t) = M^2_i(0) \left[ A_i \frac{\alpha^2_i(0) - \alpha^2_i(t)}{\alpha^2_i(0)} \right] \equiv M^2_i(0) \overline{c}_i(t), \quad i=\{1,2,3\},
   \label{eq:Ci}
\end{equation} 
with
\begin{equation}
A_i = \left\{ \frac{2}{11}, \frac{3}{2}, -\frac{8}{9} \right\},
\label{eq:coeff}
\end{equation}
and
\begin{equation}
K(t) = \frac{1}{2b_1}S_0\left( 1 - \frac{\alpha_1(t)}{\alpha_1(0)} \right),
\label{eq:Kdef}
\end{equation}
where $b_1 = 33/5$. The equivalence in equation (\ref{eq:Ci}) defines $\bar c_i(t)$. Since the squared mass-matrices of the squarks and sleptons in the gauge-eigenstate basis is diagonal for the first two generations, equations (\ref{eq:sol1}) to (\ref{eq:sol7}) represent the approximate physical masses. Again, these equations directly correspond to those of Ref.~\cite{Martin:1993ft} except for the inclusion of the non-universal sfermion contribution $K$ and minor notational differences. 

The form of equations (\ref{eq:sol1}) to (\ref{eq:sol7}) immediately allows one to write down some simple sum rules relating the running sfermion masses that are independent of the specific GUT boundary conditions. For example, 
\begin{equation}
m^2_{\tilde{u}_L} - m^2_{\tilde{d}_L} =
m^2_{\tilde{e}_L} - m^2_{\tilde{\nu}_L} = M_Z^2 (1- \sin^2 \theta_W) \cos 2 \beta ,
\label{eq:bisr1}
\end{equation}
which are the predictions of equations (3.4) and (3.5) in Ref.~\cite{Martin:1993ft}. 
Since the right hand side of this equation is rather small, this also tells us that the left handed squarks and left handed sleptons will, separately, be approximately degenerate. Two other useful sum rules are
\begin{equation}
m^2_{\tilde{u}_L} + m^2_{\tilde{d}_L}  - m^2_{\tilde{u}_R} - m^2_{\tilde{e}_R} = C_3 + 2C_2 - \frac{25}{18}C_1 \approx 4.8 M^2_{1/2}, \label{eq:Sum11}
\end{equation}
and
\begin{equation}
\frac{1}{2}\left(m^2_{\tilde{u}_L} + m^2_{\tilde{d}_L} - m^2_{\tilde{e}_L} - m^2_{\tilde{\nu}_L}\right)
+ m^2_{\tilde{d}_R} - m^2_{\tilde{e}_R}  = 2C_3 - \frac{10}{9}C_1 \approx 8.1 M^2_{1/2}.\label{eq:Sum12}
\end{equation}
The left equality in equations (\ref{eq:Sum11}) and (\ref{eq:Sum12}) are independent of the GUT scale boundary conditions and true for all values of $t$. However, the right equality is assuming the boundary condition $M_1(0)=M_2(0)=M_3(0)=M_{1/2}$ and the values for $C_i$ were obtained at a scale $Q=1\,$TeV.

\section{Boundary Conditions \label{sec:boundary_conditions}}

We will now consider the effect of fixing boundary conditions at the GUT scale according to the GUT groups $SU(5)$, $SO(10)$ and $E_6$.

\subsection{$\mathbf{SU(5)}$}

We first consider an $SU(5)$ supersymmetric GUT, breaking directly to $SU(3)\times SU(2)_L \times U(1)$ at the GUT scale, $Q_0$. Under this gauge group, all the SM fermions as well as their scalar partners are embedded in a $\mathbf{10} \oplus \mathbf{\overline{5}}$ dimensional representation, where $\tilde{L}_L$ and $\tilde{d}_R$ are in the $\mathbf{\overline{5}}$, and $\tilde{Q}_L$, $\tilde{u}_R$ and $\tilde{e}_R$ are in the $\mathbf{10}$. With this construction we do not have a universal scalar mass $m_0$ at the GUT scale, as in the cMSSM, but instead have a common $m_{\mathbf{10}}$ for the matter in the 10-plet and a common $m_{\mathbf{\overline{5}}}$ for the matter in the 5-plet. For the minimal $SU(5)$ supersymmetric GUT, the Higgs fields $H_u$ and $H_d$ belong to two distinct five-dimensional representations, $\mathbf{5}^{\prime}$ and $\overline{\mathbf{5}}^{\prime}$ respectively, so their masses at the GUT scale are unrelated. For the gaugino mass, we consider the simplest scenario, where the chiral superfields in the gauge-kinetic function are in a singlet representation of $SU(5)$ \cite{gaugino}. We then have a common gaugino mass, $M_{1/2}$, at the GUT scale. For a discussion of an $SU(5)$ GUT including $b\,$-$\tau$ Yukawa unification, see \cite{Baer:2012by}. Leaving the doublet-triplet splitting problem aside, our boundary conditions are:
\begin{eqnarray}
m_{\tilde{Q}_L}^2\left( 0 \right) &=& m_{\tilde{u}_R}^2\left( 0 \right) \;\; = \;\;   m_{\tilde{e}_R}^2\left( 0 \right)  \;\;= \;\; m_{\mathbf{10}}^2 ,\label{eq:b1}\\
m_{\tilde{d}_R}^2\left( 0 \right) &=& m_{\tilde{L}_L}^2\left( 0 \right)  \;\;= \;\; m_{\overline{\mathbf 5}}^2, \label{eq:b2}\\
m_{H_u}^2\left( 0 \right) &=&  m_{\mathbf{5}^{\prime}}^2,\label{eq:b3}\\
m_{H_d}^2\left( 0 \right )&=& m_{\overline{\mathbf{5}}^{\prime}}^2,\label{eq:b4}\\
M_1^2\left( 0 \right) &=& M_2^2\left( 0 \right)   \;\;\;\,=\;\; M_3^2\left( 0 \right)  \;\;= \;\; M_{1/2}^2.\label{eq:b5}
\end{eqnarray}
Note that inserting equations (\ref{eq:b1}) to (\ref{eq:b4}) into equation (\ref{eq:S}), we find $S_0 = m_{\mathbf{5}^{\prime}}^2 -  m_{\overline{\mathbf{5}}^{\prime}}^2 \neq 0$, so $K$, as defined by equation  (\ref{eq:K}) does not vanish at the EW scale. Considering only the sfermion sector, we have five unknowns, $m_{\overline{\mathbf{5}}}$, $m_{\mathbf{10}}$, $M_{1/2}$, $\cos 2\beta$ and $K$, and seven equations, (\ref{eq:sol1}) to (\ref{eq:sol7}), that relate these unknowns to (in principle) measurable scalar masses. If we know the EW scale mass of five sfermions, say $\tilde{u}_L$, $\tilde{d}_L$, $\tilde{e}_R$, $\tilde{u}_R$ and $\tilde{d}_R$, we have an invertible system of equations and can fully determine our five parameters.
\begin{equation}
\begin{pmatrix}
  M^2_{\tilde{u}_L}\\
  M^2_{\tilde{d}_L}\\
  M^2_{\tilde{e}_R}\\
  M^2_{\tilde{u}_R}\\
  M^2_{\tilde{d}_R}
\end{pmatrix} =
\begin{pmatrix}
0 & 1 & c_{\tilde{u}_L} & \delta_{\tilde{u}_L} & -\frac{1}{5}\\
0 & 1 & c_{\tilde{d}_L} & \delta_{\tilde{d}_L} & -\frac{1}{5}\\
0 & 1 & c_{\tilde{e}_R} & \delta_{\tilde{e}_R} & -\frac{6}{5}\\
0 & 1 & c_{\tilde{u}_R} & \delta_{\tilde{u}_R} &  \frac{4}{5}\\
1 & 0 & c_{\tilde{d}_R} & \delta_{\tilde{d}_R} & -\frac{2}{5}
\end{pmatrix}
\begin{pmatrix}
  m^2_{\overline{\mathbf{5}}}\\
  m^2_{\mathbf{10}}\\
  M^2_{1/2}\\
  \cos 2\beta\\
  K
\end{pmatrix}.
\label{eq:SU5}
\end{equation}
In this equation, and throughout the rest of the text, we have used a capital $M$ to denote the measured low energy masses, e.g.\ $M_{\tilde u_L} = m_{\tilde u_L}(M_{\tilde u_L})$. Also, we have defined
 \begin{eqnarray}
  \Delta_{\varphi} &\equiv& \delta_{\varphi}\cos 2\beta, \qquad (\varphi = \tilde u_L, \, \tilde d_L, \, \tilde e_R, \, \tilde u_R, \, \tilde d_R) \label{eq:coef1}\\
  c_{\tilde{u}_L} &\equiv& \overline{c}_3(M_{\tilde{u}_L}) + \overline{c}_2(M_{\tilde{u}_L}) + \frac{1}{36}\overline{c}_1(M_{\tilde{u}_L}),\label{eq:coef2}\\
  c_{\tilde{d}_L} &\equiv& \overline{c}_3(M_{\tilde{d}_L}) + \overline{c}_2(M_{\tilde{d}_L}) + \frac{1}{36}\overline{c}_1(M_{\tilde{d}_L}),\label{eq:coef3}\\
  c_{\tilde{e}_R} &\equiv& \overline{c}_1(M_{\tilde{e}_R}),\label{eq:coef4}\\
  c_{\tilde{u}_R} &\equiv& \overline{c}_3(M_{\tilde{u}_R}) +\frac{4}{9}\overline{c}_1(M_{\tilde{u}_R}),\label{eq:coef5}\\
 c_{\tilde{d}_R} &\equiv& \overline{c}_3(M_{\tilde{d}_R}) +\frac{1}{9}\overline{c}_1(M_{\tilde{d}_R}).\label{eq:coef6}
\end{eqnarray}
The explicit solutions determining $m_{\overline{\mathbf{5}}}$, $m_{\mathbf{10}}$, $M_{1/2}$, $\cos 2\beta$ and $K$ as function of the low energy masses are then\footnote{The results obtained for $SU(5)$ differ from those in \cite{Ananthanarayan:2004dm}, where in the expression for $\cos 2\beta$ the first term is absent.}
\begin{eqnarray}
m_{\overline{5}}^2 
&=&  \frac{1}{5X_5} 
\left[ 
(c_{\tilde{u}_L} + c_{\tilde{d}_L})(M_{\tilde{u}_R}^2 + 5M_{\tilde{d}_R}^2 - M_{\tilde{e}_R}^2 ) 
-c_{\tilde{u}_R}( M_{\tilde{u}_L}^2+ M_{\tilde{d}_L}^2 + 5M_{\tilde{d}_R}^2 - 2M_{\tilde{e}_R}^2 ) 
\right. \nonumber\\ 
&& \phantom{\frac{1}{5X_5} \Big[} \left. 
- 5c_{\tilde{d}_R}(M_{\tilde{u}_L}^2 + M_{\tilde{d}_L}^2 - M_{\tilde{u}_R}^2 - M_{\tilde{e}_R}^2 ) 
+ c_{\tilde{e}_R}(M_{\tilde{u}_L}^2 + M_{\tilde{d}_L}^2 - 2M_{\tilde{u}_R}^2- 5M_{\tilde{d}_R}^2 ) 
\right], \nonumber\\
&& \label{eq:m5}\\  
m_{10}^2 
&=& \frac{1}{5X_5} \left[ 
 (c_{\tilde{u}_L} + c_{\tilde{d}_L} )(3M_{\tilde{u}_R}^2 + 2M_{\tilde{e}_R}^2 )  
- c_{\tilde{u}_R}(3M_{\tilde{u}_L}^2+3M_{\tilde{d}_L}^2 - M_{\tilde{e}_R}^2 ) 
\right. \nonumber\\
&& \phantom{ \frac{1}{5X_5} \Big[ }
\left. 
- c_{\tilde{e}_R}(2M_{\tilde{u}_L}^2 +2M_{\tilde{d}_L}^2 +  M_{\tilde{u}_R}^2)
\right], \label{eq:m10}\\
M_{1/2}^2 
&=& \frac{1}{X_5} \left( M_{\tilde{u}_L}^2 + M_{\tilde{d}_L}^2 - M_{\tilde{u}_R}^2 - M_{\tilde{e}_R}^2 \right)~,\label{eq:m12} \\
\cos 2 \beta 
&=& \frac{1}{X_5 M_Z^2 (\sin^2\theta_W - 1)} \left[ 
 c_{\tilde{u}_L}(2M_{\tilde{d}_L}^2-M_{\tilde{u}_R}^2  - M_{\tilde{e}_R}^2 )
- c_{\tilde{d}_L}(2M_{\tilde{u}_L}^2 -  M_{\tilde{u}_R}^2 - M_{\tilde{e}_R}^2 ) 
\right. \nonumber\\
&& \phantom{\frac{1}{X_5 M_Z^2 (\sin^2\theta_W - 1)} \Big[} \left. 
+ (c_{\tilde{u}_R} + c_{\tilde{e}_R} )(M_{\tilde{u}_L}^2-M_{\tilde{d}_L}^2 )
\right], \label{eq:cos2b}
\end{eqnarray}
\begin{eqnarray}
K &=& 
\frac{1}{6X_5 (\sin^2\theta_W - 1)} \left[ 
- 3(c_{\tilde{u}_L} + c_{\tilde{d}_L})(M_{\tilde{u}_R}^2 - M_{\tilde{e}_R}^2 ) 
+ 3c_{\tilde{u}_R}(M_{\tilde{u}_L}^2 + M_{\tilde{d}_L}^2 - 2M_{\tilde{e}_R}^2 ) 
\right. \nonumber\\
&&  \hspace*{32mm}
- 3c_{\tilde{e}_R}( M_{\tilde{u}_L}^2 + M_{\tilde{d}_L}^2 - 2M_{\tilde{u}_R}^2) 
\nonumber \\
&& \left.
+ 2\sin^2\theta_W\left( 
            c_{\tilde{u}_L}(4M_{\tilde{u}_R}^2-5M_{\tilde{d}_L}^2 + M_{\tilde{e}_R}^2 ) 
           + c_{\tilde{d}_L}( 5M_{\tilde{u}_L}^2 - M_{\tilde{u}_R}^2- 4M_{\tilde{e}_R}^2  )
\right. \right. \nonumber\\
&&\left.\left.  \hspace*{18mm}
            -c_{\tilde{u}_R}( 4M_{\tilde{u}_L}^2 - M_{\tilde{d}_L}^2 - 3M_{\tilde{e}_R}^2 ) 
           -c_{\tilde{e}_R}( M_{\tilde{u}_L}^2 - 4M_{\tilde{d}_L}^2  + 3M_{\tilde{u}_R}^2) 
    \right)  \right],\label{eq:K}
\end{eqnarray}
where $X_5$ is given by:
\begin{equation}
X_5 = c_{\tilde{u}_L} + c_{\tilde{d}_L}  - c_{\tilde{u}_R} - c_{\tilde{e}_R}
\label{eq:X5}
\end{equation}

We had seven equations, (\ref{eq:sol1}) to (\ref{eq:sol7}), but only five unknowns, so we should have two constraints left over. These are provided by the  sum rules. The unused equations are (\ref{eq:sol5}) and (\ref{eq:sol6}); their difference provides the second equality of equation (\ref{eq:bisr1}) while their sum is part of equation (\ref{eq:Sum12}), where it has been combined with other masses to remove the non-$C_i$ terms. The other sum rule, equation (\ref{eq:Sum11}), is just a re-expression of equation (\ref{eq:m12}).

Some simplification of equations  (\ref{eq:sol1}) to (\ref{eq:sol7}) is possible by allowing more approximations. For example, since the running of the gauge couplings is logarithmic, $\bar c_i$ only have a rather small dependence on the scale where they are evaluated, and so $c_{\tilde u_L} \approx c_{\tilde d_L}$. Also $\bar c_1$ is numerically rather small and its contribution is diminished by its small coefficients in equations (\ref{eq:coef5} - \ref{eq:coef6}), so  $c_{\tilde u_R} \approx c_{\tilde d_R}$ and $c_{\tilde e_R}$ can be neglected. Furthermore, for TeV scale sfermions, the contribution from the electroweak D-term is small, since it is added in quadrature, allowing one to neglect $\delta_\varphi$. Finally, evaluating the masses at a common scale one finds
\begin{eqnarray}
m_{\overline{5}}^2 
&\approx&  \frac{1}{5X_5} 
\left[ 
2c_L(m_{\tilde{u}_R}^2 + 5m_{\tilde{d}_R}^2 - m_{\tilde{e}_R}^2 ) 
-c_R( 12m_{\tilde{u}_L}^2 + 5m_{\tilde{d}_R}^2
  - 5m_{\tilde{u}_R}^2 - 7M_{\tilde{e}_R}^2 ) 
\right], \nonumber\\
&& \label{eq:m5app}\\  
m_{10}^2 
&\approx& \frac{1}{5X_5} \left[ 
 2c_L(3m_{\tilde{u}_R}^2 + 2m_{\tilde{e}_R}^2 )  
- c_R(6m_{\tilde{u}_L}^2 - m_{\tilde{e}_R}^2 ) 
\right],  \label{eq:m10app}\\
M_{1/2}^2 
&\approx& \frac{1}{X_5} \left( 2m_{\tilde{u}_L}^2  - m_{\tilde{u}_R}^2 - m_{\tilde{e}_R}^2 \right),\label{eq:m12app} \\
K &\approx& \frac{1}{X_5} \left[ c_L \left( m_{\tilde{u}_R}^2  - m_{\tilde{e}_R}^2 \right) - c_R \left( m_{\tilde{u}_L}^2  - m_{\tilde{e}_R}^2 \right) \right],\label{eq:Kapp}
\end{eqnarray}
and $X_5$ takes the simplified form
\begin{equation}
X_5 = 2c_L-c_R.
\end{equation}
In an obvious notation, $c_L \equiv c_{\tilde{u}_L} \approx c_{\tilde{d}_L}$ and $c_R \equiv c_{\tilde{u}_R} \approx c_{\tilde{d}_R}$. The equation for $\cos 2 \beta$ has dropped out of these approximate equations since the electroweak D-term has been neglected. 

\subsection{$\mathbf{SO(10)}$}
 
We now consider grand unification with boundary conditions of $SO(10)$. Now all the squarks and sleptons are embedded in the fundamental 16-dimensional irrep of $SO(10)$, including the right-handed sneutrino. We shall consider here the breaking scenario
\begin{equation}
SO(10) \rightarrow SU(5) \otimes U(1)_x \rightarrow SU(3) \otimes SU(2)_L \otimes U(1),
\label{eq:Chain}
\end{equation}
where we assume that the intermediate breakings all occur around the GUT scale, motivated by the successful unification of the gauge couplings in the MSSM. It is important to note that $SO(10)$ is a gauge group of rank-5, which means that the breaking chain (\ref{eq:Chain}) involves the reduction of rank from 5 to 4. In general, if one considers a supersymmetric model with $n$ extra $U(1)$s and assume a Higgs type mechanism, the extra $U(1)$s may be spontaneously broken by the vevs of the scalar components of the Higgs superfields $\Phi$ and $\overline{\Phi}$, with charges $Q_{k\Phi}$ and $-Q_{k\Phi}$ respectively. The scalar supersymmetric potential with D-terms  included is
 \begin{equation}
V_{SUSY} = \frac{1}{M^{4n-6}}\left( \lvert \Phi \rvert^2  + \lvert \overline{\Phi} \rvert^2 \right)\lvert \Phi \overline{\Phi} \rvert^{2n-2} + \sum_k \frac{g_k^2}{2} \left(Q_{k\Phi} \left( \lvert \Phi \rvert^2  - \lvert \overline{\Phi} \rvert^2 \right)  + \sum_a Q_{ka} \lvert \varphi_a \rvert^2 \right)^2
\label{eq:D1}
\end{equation}
and the additional soft SUSY breaking terms have the form \cite{Kolda:1995iw}
 \begin{equation}
V_{\rm soft} = m_{\Phi}^2 \lvert \Phi \rvert^2 + m_{\overline{\Phi}}^2 \lvert \overline{\Phi} \rvert^2,
\label{eq:soft}
\end{equation}
where $ \varphi_a$ plays the role of the usual MSSM scalar fields, $g_k$ are the diverse $U(1)_k$ gauge couplings, $ m_{\Phi}^2$ and $m_{\overline{\Phi}}^2$ are soft scalar masses and $M$ is a mass of order the Planck scale. The scalar potential is assumed to receive a non-trivial vev in a nearly D-flat direction of the form
 \begin{equation}
\langle \Phi \rangle^2 \approx \langle \overline{\Phi} \rangle^2 \approx \left[ \frac{-\left( m_{\overline{\Phi}}^2 + m_{\Phi}^2 \right) M^{4n-6}}{4n -2} \right]^{1/(2n-2)},
\label{eq:vevs}
\end{equation}
where $m_{\overline{\Phi}}^2 + m_{\Phi}^2$ must be negative at the scale of $\langle \Phi \rangle$. After integrating out the superfields  $\Phi$ and $\overline{\Phi}$, the corrections to the soft scalar masses for the surviving fields $\varphi_a$ are proportional to their charges under the broken $U(1)_k$, having the form
 \begin{equation}
\Delta m_a^2 = \sum_k Q_{ka} g_k^2 D_k,
\label{eq:correction}
\end{equation}
where the D-term is given by\footnote{In principle, the form of the D-terms can be rather more complicated, reflecting non-trivial features of the breaking mechanism. Usually, one considers $D_k$ to be a parameter of our ignorance of these details.}
 \begin{equation}
D_k = \frac{\frac{1}{2} \left(m_{\overline{\Phi}}^2 - m_{\Phi}^2 \right) Q_{k\Phi} }{\displaystyle\sum_l g_l^2 Q_{l\Phi}^2}.
\label{eq:d}
\end{equation}
One can see that the D-terms depend only on the soft masses $m_{\Phi}$, $m_{\overline{\Phi}}$ and on the $U(1)_k$ charges, and not on the form of the scalar potential (\ref{eq:D1}) itself. Even if the scale of spontaneous symmetry breaking governed by equation (\ref{eq:vevs}) is well above $m^2_{\rm soft}$, the D-term contributions will remain of order the square of the soft scalar masses.

With this in mind, the breaking of the additional $U(1)_x$ in the chain of equation (\ref{eq:Chain}) at the high scale will involve a D-term contribution of order $m_{\rm soft}^2$. For the Higgs sector, we will consider a simple scenario where both the up-type and down-type Higgs fields are embedded in a 10-dimensional irrep of $SO(10)$. Then we have a common scalar mass $m_{\mathbf{16}}$ for the sfermions at the GUT scale, and a common mass $m_{\mathbf{10}}$ for the Higgs fields. Additionally, due to rank reduction after the breaking of $SO(10)$, one has D-term contributions of the form of equation (\ref{eq:correction}). For the gaugino masses, the argument that justifies the common GUT scale mass $M_{1/2}$ for $SU(5)$ remains valid. We have then the following boundary conditions:
\begin{eqnarray}
m_{\tilde{Q}_L}^2\left( 0 \right) \;\;=\;\; m_{\tilde{u}_R}^2\left(0 \right) \;\;=\;\; \,  m_{\tilde{e}_R}^2\left( 0 \right) &=& m_{\mathbf{16}}^2 + g_{10}^2D, \label{eq:bc1}\\
m_{\tilde{d}_R}^2\left( 0 \right) \;\;=\;\; m_{\tilde{L}_L}^2\left( 0 \right) &=& m_{\mathbf{16}}^2 - 3g_{10}^2D, \label{eq:bc2}\\
m_{H_u}^2\left( 0 \right) &=& m_{\mathbf{10}^2} - 2g_{10}^2D,\label{eq:bc3}\\
m_{H_d}^2\left( 0 \right) &=& m_{\mathbf{10}^2} + 2g_{10}^2D,\label{eq:bc4}\\
M_1\left( 0 \right) \;\;=\;\;\; M_2\left( 0 \right) \;\;\,=\;\;\;\; M_3\left( 0 \right) &=& M_{1/2},\label{eq:bc5}
\end{eqnarray}
where $g_{10}$ is the common value of the gauge couplings at the GUT scale. 

One interesting difference between this scenario and $SU(5)$ unification is the extra relation between Higgs masses at the GUT scale, which, when inserted into equation (\ref{eq:S}), results in
\begin{equation}
S_0 = -4g^2_{10}D .
\label{eq:StGSO10}
\end{equation}
As before, considering only the sfermion sector we have five unknowns, $m_{\mathbf{16}}$, $g^2_{10}D$, $M_{1/2}$, $\cos 2\beta$ and $K$, with seven equations. The measurement of $M_{\tilde{u}_L}$, $M_{\tilde{d}_L}$, $M_{\tilde{e}_R}$, $M_{\tilde{u}_R}$ and $M_{\tilde{d}_R}$ is sufficient to determine these five parameters using the invertible system
\begin{equation}
\begin{pmatrix}
  M^2_{\tilde{u}_L}\\
  M^2_{\tilde{d}_L}\\
  M^2_{\tilde{e}_R}\\
  M^2_{\tilde{u}_R}\\
  M^2_{\tilde{d}_R}
\end{pmatrix} =
\begin{pmatrix}
1 & 1 & c_{\tilde{u}_L} & \delta_{\tilde{u}_L} & -\frac{1}{5}\\
1 & 1 & c_{\tilde{d}_L} & \delta_{\tilde{d}_L} & -\frac{1}{5}\\
1 & 1 & c_{\tilde{e}_R} & \delta_{\tilde{e}_R} & -\frac{6}{5}\\
1 & 1 & c_{\tilde{u}_R} & \delta_{\tilde{u}_R} &  \frac{4}{5}\\
1 & -3 & c_{\tilde{d}_R} & \delta_{\tilde{d}_R} & -\frac{2}{5}
\end{pmatrix}
\begin{pmatrix}
  m^2_{\mathbf{16}}\\
  g^2_{10}D\\
  M^2_{1/2}\\
  \cos 2\beta\\
  K
\end{pmatrix} .
\label{eq:SO10}
\end{equation}
Here, $M_{1/2}$ and $\cos 2\beta$ and $K$ are given by the same expressions as for $SU(5)$, equations~(\ref{eq:m12} - \ref{eq:K}), whereas $m_{\mathbf{16}}$ and $g^2_{10}D$ are given by\footnote{The result obtained for $m_{\mathbf{16}}$ differ from that in \cite{Ananthanarayan:2004dm}, in the term proportional to $c_{\tilde{d}_R}$.}
\begin{eqnarray}
m_{16}^2 &=& \frac{1}{4X_5} \left[ -c_{\tilde{u}_R}(2M_{\tilde{d}_L}^2 + M_{\tilde{d}_R}^2 - M_{\tilde{e}_R}^2 + 2M_{\tilde{u}_L}^2) - c_{\tilde{e}_R}(M_{\tilde{d}_L}^2 + M_{\tilde{d}_R}^2 + M_{\tilde{u}_L}^2 + M_{\tilde{u}_R}^2)\right.\nonumber\\ 
&& \left. + (c_{\tilde{d}_L} + c_{\tilde{u}_L})(M_{\tilde{d}_R}^2 + M_{\tilde{e}_R}^2 + 2M_{\tilde{u}_R}^2) + c_{\tilde{d}_R}(-M_{\tilde{d}_L}^2 + M_{\tilde{e}_R}^2 - M_{\tilde{u}_L}^2 + M_{\tilde{u}_R}^2) \right], \label{eq:m16}\\
g_{10}^2D &=& \frac{1}{20X_5} \left[ -c_{\tilde{u}_R}(2M_{\tilde{d}_L}^2 - 5M_{\tilde{d}_R}^2 + M_{\tilde{e}_R}^2 + 2M_{\tilde{u}_L}^2) + c_{\tilde{e}_R}(-3M_{\tilde{d}_L}^2 + 5M_{\tilde{d}_R}^2 - 3M_{\tilde{u}_L}^2 + M_{\tilde{u}_R}^2)\right.\nonumber\\ 
&& \left. - (c_{\tilde{d}_L} + c_{\tilde{u}_L})(5M_{\tilde{d}_R}^2 - 3M_{\tilde{e}_R}^2 - 2M_{\tilde{u}_R}^2) + 5c_{\tilde{d}_R}(M_{\tilde{d}_L}^2 - M_{\tilde{e}_R}^2 + M_{\tilde{u}_L}^2 - M_{\tilde{u}_R}^2) \right].\label{eq:D10}
\end{eqnarray}
The choice of the Higgs fields in a 10-plet enables us to relate their GUT scale masses through the relation (\ref{eq:StGSO10}). If we plug this expression into equation (\ref{eq:K}), we obtain
\begin{equation}
K(t) = \frac{-4g^2_{10}D}{2b_1}\left( 1 - \frac{\alpha_1(t)}{\alpha_1(0)} \right).
\label{eq:constraint}
\end{equation}
Using the expressions for $K$, equation (\ref{eq:K}), and for $g^2_{10}D$, equation (\ref{eq:D10}), which are explicitly dependent on the low energy squark and slepton masses, we have a further constraint upon the sfermion masses. 

This new relation is useful in distinguishing between GUT groups since it provides a direct constraint involving only the sfermion masses. If we do indeed find a first (and/or second) generation of sfermions at the LHC, measuring four of these masses will provide an $SO(10)$ prediction of the fifth. To see the significance of this, suppose that we find a first or second generation of sfermions, and measure the five masses $M_{\tilde{u}_L}$, $M_{\tilde{d}_L}$, $M_{\tilde{e}_R}$, $M_{\tilde{u}_R}$ and $M_{\tilde{d}_R}$. We cannot yet use equations (\ref{eq:m5} - \ref{eq:K}) or (\ref{eq:m16} - \ref{eq:D10}) to determine the model parameters since we do not yet know which boundary conditions to apply. However, after inserting the expressions for $K$ and $g_{10}^2D$ found in equations (\ref{eq:K}) and (\ref{eq:D10}) respectively, equation (\ref{eq:constraint}) provides an $SO(10)$ prediction of the $M_{\tilde{d}_R}$ which we can compare to the measured value. One can see an example of this in Table~\ref{table:mass} (lower section), where we have presented three scenarios, whose details we will use for a numerical comparison with SOFTSUSY in Section~\ref{sec:softsusy}.

In this table, values of the masses $M_{\tilde{u}_L}$, $M_{\tilde{d}_L}$, $M_{\tilde{e}_R}$ and $M_{\tilde{u}_R}$ have been chosen, consistent with unification and $ -1 < \cos 2\beta < 0$. For $SU(5)$ we have no constraint on the value of $M_{\tilde{d}_R}$ so must also treat this as an input, but for $SO(10)$, expression (\ref{eq:constraint}) fixes the value of $M_{\tilde{d}_R}$ as shown. Also note that some choices for the masses $M_{\tilde{u}_L}$, $M_{\tilde{d}_L}$, $M_{\tilde{e}_R}$ and $M_{\tilde{u}_R}$, that are acceptable for $SU(5)$ and for which a equation (\ref{eq:constraint}) provides a seemingly reasonable solution for $M_{\tilde{r}_R}$ in $SO(10)$, may actually be forbidden for $SO(10)$ since $m_{\mathbf{16}}^2 < 3g_{10}^2D$ and thus $m_{\tilde d_R}^2(0)<0$ (though this is not the case for any of the scenarios shown).
\begin{table}[ht]
\begin{center}
\begin{tabular}{|c|c|c|c|c|}
\hline
\rule[-1mm]{0mm}{6mm} && ~~Scenario 1~~ & ~~Scenario 2~~ & ~~Scenario 3~~ \\
\hline
\hline
\rule[-2mm]{0mm}{7mm}
&$m_{\overline{\mathbf{5}}}$   & 781.7 & 893.7 & 2856.6  \\
\cline{2-5}
\rule[-2mm]{0mm}{7mm}
$SU(5)$ &$m_{\mathbf{10}}$  & 654.8 & 1385.0 & 2690.5 \\
\cline{2-5}
\rule[-2mm]{0mm}{7mm}
&$m_{\overline{\mathbf{5}}^\prime}$ & 800 & 1800 & 2700 \\
\hline 
\rule[-2mm]{0mm}{7mm}
&$m_{\mathbf{16}}$ & 669.9 & 1268.9 & 2811.6 \\
\cline{2-5}
\rule[-2mm]{0mm}{7mm}
$SO(10)$&$m_{\mathbf{10}}$  & 800 & 1800 & 2700 \\
\cline{2-5}
\rule[-2mm]{0mm}{7mm}
&$g^2_{10}D$  & -19.971 $\times \mathrm{10^3}$ &308.263 $\times \mathrm{10^3}$ & -666.100 $\times \mathrm{10^3}$\\
\hline 
\rule[-2mm]{0mm}{7mm}
$SU(5)$&$S_0$ & 79.886 $\times \mathrm{10^3}$  & -1233.05  $\times \mathrm{10^3}$ &  2664.40 $\times \mathrm{10^3}$ \\
\cline{2-5}
\rule[-2mm]{0mm}{7mm}
\& $SO(10)$ &$\tan \beta$ & 6.1 & 8.0 & 4.6 \\
\hline
\hline
\rule[-2mm]{0mm}{7mm}
&$M_{\tilde{u}_L}$ & 1550 & 1951 & 3550 \\
\cline{2-5}
\rule[-2mm]{0mm}{7mm}
$SU(5)$&$M_{\tilde{d}_L}$ & 1552 & 1953 & 3551 \\
\cline{2-5}
\rule[-2mm]{0mm}{7mm}
\& $SO(10)$ &$M_{\tilde{e}_R}$ & 700 & 1430 & 2700 \\
\cline{2-5}
\rule[-2mm]{0mm}{7mm}
&$M_{\tilde{u}_R}$ & 1500 & 1898 & 3500 \\
\hline
\rule[-2mm]{0mm}{7mm}
$SU(5)$&$M_{\tilde{d}_R}$ & 1550 & 1600 & 3600 \\
\hline
\rule[-2mm]{0mm}{7mm} 
$SO(10)$&$M_{\tilde{d}_R}$ &  1518 &  1566 &  3830 \\
\hline
\end{tabular}
\caption{Example scenarios to demonstrate the use of the additional $SO(10)$ sum rule and test the sum rules with SOFTSUSY. All masses are GeV (though $S_0$ and $g^2_{10}D$ have dimension mass$^2$).}
\label{table:mass}
\end{center}
\end{table}

As  mentioned earlier, some caution is required, since this additional sum rule is characteristic of the choosing the Higgs fields to be in the $\mathbf{10}$ of $SO(10)$. It would be interesting to perform further studies to investigate which constraints on the masses would arise with Higgs embedded in a $\mathbf{120}$, or a $\mathbf{126}$, or even combinations of them.

\subsection{$E_6$} \label{subsec:e6}

For unification under the group $E_6$, the fundamental sfermions and Higgs are embedded in a $\mathbf{27}$ irrep together with additional exotic matter. For now, let us consider a simple scenario where all the extra fields (i.e.\ those that don't appear in the cMSSM) are integrated out at the high scale and where the intermediate breaking of $E_6$ subgroups all occur around the GUT scale. Our motivation is to explore further constraints on the squark and slepton masses due to placing all our matter in a {\bf 27}-plet with a common scalar mass $m_{ \bf 27}$ at the GUT scale with GUT scale masses separated only by D-terms.

We consider the breaking
\begin{equation}
E_6 \rightarrow SO(10) \otimes U(1)_S \rightarrow SU(5) \otimes U(1)_S \otimes U(1)_X \rightarrow SU(3) \otimes SU(2)_L \otimes U(1).
\label{eq:E6Chain}
\end{equation}
$E_6$ is a rank-6 group, so the breaking to the SM group involves a rank reduction of two units and we have two D-term contributions from the breaking of $U(1)_S$ and $U(1)_X$ at the high scale, where the common gauge coupling has the value $g_6^2$. As for $SU(5)$ and $SO(10)$, we assume a common value $M_{1/2}$ for the gaugino masses at the high scale. The boundary conditions are then:
\begin{eqnarray}
m_{\tilde{Q}_L}^2\left( 0 \right) \;\;=\;\; m_{\tilde{u}_R}^2\left(0 \right) \;\;=\;\; \,  m_{\tilde{e}_R}^2\left( 0 \right) &=&  m^2_{\mathbf{27}} - g^2_{6}D_S + g^2_{6}D_X, \label{eq:e6bc1}\\
m_{\tilde{d}_R}^2\left( 0 \right) \;\;=\;\; m_{\tilde{L}_L}^2\left( 0 \right) &=& m^2_{\mathbf{27}} - g^2_{6}D_S - 3g^2_{6}D_X, \label{eq:e6bc2}\\
m_{H_u}^2\left( 0 \right) &=& m^2_{\mathbf{27}} + 2g^2_{6}D_S - 2g^2_{6}D_X,\label{eq:e6bc3}\\
m_{H_d}^2\left( 0 \right) &=& m^2_{\mathbf{27}} + 2g^2_{6}D_S + 2g^2_{6}D_X,\label{eq:e6bc4}\\
M_1\left( 0 \right) \;\;=\;\;\; M_2\left( 0 \right) \;\;\,=\;\;\;\; M_3\left( 0 \right) &=& M_{1/2},\label{eq:e6bc5}
\end{eqnarray}
where at the GUT scale we have
\begin{equation}
S_0 = -4g^2_{6}D_X~.
\label{eq:StGE6}
\end{equation}
We have six unknowns, $m_{\mathbf{27}}$, $g^2_{6}D_S$, $g^2_{6}D_X$, $M_{1/2}$, $\cos 2\beta$ and $K$, with seven equations. However, all the sfermions have the same $U(1)_S$ charge, so $m^2_{\mathbf{27}}$ and $g^2_{6}D_S$ always appears in the combination $m^2_{\mathbf{27}} - g^2_{6}D_S$ in the sfermion boundary conditions, and cannot be disentangled without extra input from the Higgs sector. Given that we assume $E_6$ breaks to $SO(10) \otimes U(1)_S$ we may identify $m^2_{\mathbf{16}}$ with $m^2_{\mathbf{27}} - g^2_{6}D_S$ and $m^2_{\mathbf{10}}$ with $m^2_{\mathbf{27}} + 2g^2_{6}D_S$. Then the previous equations for $SO(10)$, equations~(\ref{eq:m12} - \ref{eq:K}) and (\ref{eq:m16} - \ref{eq:D10}), apply with $m^2_{\mathbf{16}}$ replaced by $m^2_{\mathbf{27}} - g^2_{6}D_S$. The analysis is then reduced to that of $SO(10)$.

\section{Including Additional Matter: The E$_6$SSM}
\label{sec:e6ssm}

In Section~\ref{sec:boundary_conditions} we demonstrated that one may determine some of the free parameters of a grand unified model just by the measurement of the sfermion masses. For $SU(5)$ and $SO(10)$ we found analytic solutions for those parameters and additional constraints on the squark and slepton masses of the first two generations. In Subsection~\ref{subsec:e6} we considered the GUT group $E_6$ and found that it is not possible to determine all the boundary condition parameters of the sfermion sector from the sfermion masses alone, since one could not disentangle the {\bf 27}-plet mass from the $U(1)_S$ D-term. The analysis of the mass spectrum reduced to that of $SO(10)$ with an effective $m_{\mathbf{16}}$. 

However, this $E_6$ analysis was done with the assumption that the extra matter that fills up the {\bf 27}-plet remains at the high scale, so that the RGEs remain as they were for $SU(5)$ and $SO(10)$. In principle, there is no reason why this additional matter should not be present at low energy scales, as described by the Exceptional Supersymmetric Standard Model (E$_6$SSM) \cite{e6ssm}. This model is inspired by the breaking of a GUT symmetry $E_6$ down to the gauge group of the SM with an additional $U(1)_N$. The particular choice of $U(1)_N$ remaining at low energies is such that only the right-handed neutrino is left neutral allowing it to naturally maintain a high mass, thereby facilitating the see-saw mechanism for neutrino masses. 

In order to preserve gauge coupling unification, two additional $SU(2)_L$ doublets, $H^{\prime}$ and $\overline{H}^{\prime}$ are required. These are presumed to arise from incomplete $\mathbf{27^{\prime}}$ and $\mathbf{\overline{27}^{\prime}}$ irreps respectively and lead to a doublet-25plet splitting similar to the doublet-triplet splitting problem that we can find both in $SU(5)$ and $SO(10)$ GUTs. These additional fields are also useful in providing a solution to the baryon asymmetry problem~\cite{King:2008qb}. The Higgs fields are now embedded in the {\bf 27}-plet, so three generations of Higgs are required (as well as an additional singlet Higgs for each generation), though only the third generation Higgs gain vevs. This latter requirement is arranged using an approximate $Z_2^H$ symmetry, and additional $Z_2^B$ or $Z_2^L$ symmetries (analogous to R-parity) may be invoked to prevent Flavour Changing Neutral Currents. The theory gives rise to a distinctive low energy spectrum~\cite{Athron:2009bs,e6ssm_pheno}, which includes colour triplet fermions that may be discovered at the LHC. In \cite{e6ssm,Athron:2009bs} these colour triplet fermions are labelled $D$ and $\bar D$ but here, since we have so many $D$s already, we shall refer to them as $T$ and $\bar T$ (where ``$T$'' stands for {\it triplet}).

To perform an analysis of the first and/or second generation sfermion sector, along the lines of our analysis of $SU(5)$ and $SO(10)$, we must take into account the contribution of the extra fields, and the extra $U(1)_N$ symmetry,  to the RGEs.  In particular we will have an extra $S^{\prime}$ contribution from the extra $U(1)_N$, a D-term from the breaking of $U(1)_N$ at the TeV scale, $g_1^{\prime 2} D^{\prime}$, analougous to the electroweak $\Delta_{\phi}$ and a high scale D-term $g_6^2D$ arrising from the breaking of the additional $U(1)$ combination orthogonal to $U(1)_N$, which we shall refer to as $U(1)_M$. The charges of the fields in the $\mathbf{27}$ with respect to $U(1)_N$ and $U(1)_M$ are given in Table (\ref{table:charges}).
\begin{table}[ht]
\begin{center}
\begin{tabular}{|c|c|c|c|c|c|c|c|c|c|c|c|}
\hline 
&\lefteqn{\phantom{\Big(}} $Q_L$ & $u_R$ & $d_R$ & $L_L$ & $e_R$ & $N_R$ & $S$ & $H_2$ & $H_1$ & $T$ & $\overline{T}$ \\
\hline
$\lefteqn{\phantom{\Big(}} \sqrt{40}Q_N$ & 1 & 1 & 2 & 2 & 1 & 0 & 5 & -2 & -3 & -2 & -3 \\
\hline
$\lefteqn{\phantom{\Big(}} \sqrt{\frac{200}{3}}Q_M$ & 1 & 1 & -2 & -2 & 1 & 4 & 1 & -2 & 1 & -2 & 1  \\
\hline
\end{tabular}
\end{center}
\caption[]{\textit{ $U(1)_N$ and $U(1)_M$ normalized charges of the fields in the ${\mathbf{27}}$ of $E_6$}}
\label{table:charges}
\end{table}

One finds that the RGEs for $S$ and $S^\prime$ are coupled, 
\begin{eqnarray}
\frac{dS}{dt} &=& \frac{96}{5} \frac{\alpha_1}{4 \pi}S - \frac{1}{5} \frac{\alpha^{\prime}_1 }{4 \pi} S^{\prime}, \label{eq:SE6_1}\\
\frac{dS^{\prime}}{dt} &=& -\frac{24}{5} \frac{\alpha_1}{4 \pi}S + \frac{94}{5} \frac{\alpha^{\prime}_1 }{4 \pi}S^{\prime},
\label{eq:SE6_2}
\end{eqnarray}
so a simple analytical expression of the form of equation (\ref{eq:Sevolution}), as one had for $SU(5)$ and $SO(10)$, is not available. Since most of the $E_6$ matter is now in a single multiplet, their contributions to $S$ cancel, leaving only the contributions from $H^{\prime}$ and $\overline{H}^{\prime}$, giving
\begin{eqnarray}
S_0 \:  \:  \: \equiv  \:  \: \: S(0) &=& -m^2_{27^{\prime}} + m^2_{\overline{27}^{\prime}}, \label{eq:StGE6_1}\\
S^\prime_0 \:  \:  \: \equiv  \:  \: S^{\prime}(0) &=& 4m^2_{27^{\prime}} - 4m^2_{\overline{27}^{\prime}}.
\label{eq:StGE6_2}
\end{eqnarray}
Therefore in scenarios with unified $H^{\prime}$ and $\overline{H}^{\prime}$ masses, the $S(t)$ and $S^{\prime}(t)$ terms will be identically zero for all scales. Integrating (\ref{eq:SE6_1}) and (\ref{eq:SE6_2}) we get the coupled equations,
\begin{eqnarray}
S(t) &=& S_0 + \frac{1}{5}K^{\prime}(t) - \frac{96}{5}K(t) , \label{eq:Sevol_1}\\
S^{\prime}(t) &=& -\frac{1}{4}S_0 - \frac{94}{5}K^{\prime}(t) + \frac{24}{5}K(t) ,
\label{eq:Sevol_2}
\end{eqnarray}
where we have used $S_0^\prime = -4S_0$. Here $K^\prime$ is the $U(1)_N$ equivalent of $K$ with a definition analogous to equation (\ref{eq:K}).

The integrated RGEs are now,
\begin{eqnarray}
m^2_{\tilde{u}_L}(t) &=& m^2_{\tilde{Q}_L}(0) + C^{E_6}_3 + C^{E_6}_2 + \frac{1}{36} C^{E_6}_1 + \frac{1}{4} C^{\prime}_1 + \Delta_{u_L}+ \Delta^{\prime}_{u_L} - \frac{1}{5}K - \frac{1}{20}K^{\prime}  - g_6^2 D, \hspace{8mm} \label{eq:essm1}\\
m^2_{\tilde{d}_L}(t) &=& m^2_{\tilde{Q}_L}(0) + C^{E_6}_3 + C^{E_6}_2 + \frac{1}{36} C^{E_6}_1 + \frac{1}{4} C^{\prime}_1 + \Delta_{d_L}+ \Delta^{\prime}_{d_L} - \frac{1}{5}K- \frac{1}{20}K^{\prime}  - g_6^2 D,\label{eq:essm2}\\
m^2_{\tilde{u}_R}(t) &=& m^2_{\tilde{u}_R}(0) + C^{E_6}_3 + \frac{4}{9} C^{E_6}_1 + \frac{1}{4} C^{\prime}_1 + \Delta_{u_R}+ \Delta^{\prime}_{u_R} + \frac{4}{5}K - \frac{1}{20}K^{\prime}  - g_6^2 D,\label{eq:essm3}\\
m^2_{\tilde{d}_R}(t) &=& m^2_{\tilde{d}_R}(0) + C^{E_6}_3 + \frac{1}{9} C^{E_6}_1 + C^{\prime}_1 + \Delta_{d_R}+ \Delta^{\prime}_{d_R} - \frac{2}{5}K - \frac{1}{10}K^{\prime}  + 2 g_6^2 D,\label{eq:essm4}\\
m^2_{\tilde{e}_L}(t) &=& m^2_{\tilde{L}_L}(0) + C^{E_6}_2 + \frac{1}{4} C^{E_6}_1 + C^{\prime}_1 + \Delta_{e_L}+ \Delta^{\prime}_{e_L} + \frac{3}{5}K - \frac{1}{10}K^{\prime}  + 2 g_6^2 D,\label{eq:essm5}\\
m^2_{\tilde{\nu}_L}(t) &=& m^2_{\tilde{L}_L}(0) + C^{E_6}_2 + \frac{1}{4} C^{E_6}_1 + C^{\prime}_1 + \Delta_{\nu_L}+ \Delta^{\prime}_{\nu_L} + \frac{3}{5}K - \frac{1}{10}K^{\prime}  + 2 g_6^2 D,\label{eq:essm6}\\
m^2_{\tilde{e}_R}(t) &=& m^2_{\tilde{e}_R}(0) + C^{E_6}_1 + C^{\prime}_1 + \Delta_{e_R}+ \Delta^{\prime}_{e_R} - \frac{6}{5}K - \frac{1}{20}K^{\prime}  - g_6^2 D,\label{eq:essm7}
\end{eqnarray}
where
\begin{equation}
   C^{E_6}_i(t) = M^2_i(0) \left[ A^{E_6}_i \frac{\alpha^2_i(0) - \alpha^2_i(t)}{\alpha^2_i(0)} \right] = M^2_i(0) \overline{c}^{E_6}_i(t)~,i=\{1,2,3,4\}~,
   \label{eq:CiE6}
\end{equation} 
with
\begin{equation}
A^{E_6}_i = \left\{ \frac{1}{8}, \frac{3}{8}, \frac{20}{3}, \frac{1}{47} \right\}~.
\label{eq:coeffE6}
\end{equation}
Note that here we have identified $C^{E_6}_4 \equiv C^{\prime}_1$, and $M_4$ as the mass of the $U(1)_N$ gaugino. Also, the $U(1)_N$ D-term is 
\begin{equation}
\Delta^\prime_{\varphi} = \frac{g_1^{\prime \, 2}}{2 \sqrt{40}}  Q^N_\varphi D^\prime,
\end{equation}
where we define
\begin{equation}
D^{\prime} \equiv \sqrt{40}\left(  Q^{N}_{H_1} v^2_d + Q^N_{H_2} v^2_u + Q^N_S v^2_s \right),
   \label{eq:D}
\end{equation} 
with $Q^{N}_{\varphi}$ the $U(1)_N$ charges of the field $\varphi$ and $v_{d,u,s}$ the down-type, up-type and singlet Higgs vevs respectively. In principle this $D^\prime$ is entirely measurable at low energies from the Higgs properties and $Z^\prime$ mass, but this will be very challenging and we will here assume that $D^\prime$ is an unknown.

Inserting the $U(1)_N$ charges into $\Delta^\prime_\varphi$ in equations (\ref{eq:essm1}) to (\ref{eq:essm7}), we notice that the $K^{\prime}$ and the $g^{\prime 2}_1D^{\prime}$ always appear in the combination 
\begin{equation}
20 \, D_N \equiv \frac{1}{4} g^{\prime 2}_1D^{\prime} -K^{\prime},\label{eq:DN}
\end{equation}
so cannot be disentangled without extra information (the factor $20$ is for later notational convenience).

 We have six unknowns and seven equations so this time we must make use of either $m^2_{\tilde{e}_L}(t)$ or $m^2_{\tilde{\nu}_L}(t)$. Unfortunately, neither is a good choice since they fail to provide orthogonal information on the system, preventing us from determining all six parameters. To overcome this, it may be possible to also consider the first and second generation exotic colored triplet fileds, $T_{1,2}$ or $\overline{T_{1,2}}$, or more precisely their scalar partners. In order to provide analytic solutions, as our previous treatment,  we require small Yukawa couplings, $\kappa_{1,2}$. Further discussion of these fields can be found in Ref.~\cite{Athron:2009bs}. If $\kappa_{1,2}$ are small we have an extra equation for the $\tilde{T}_{1,2}$ mass,
\begin{equation}
m^2_{\tilde{T}_{1,2}}(t) = m^2_{\tilde{T}_{1,2}}(0) + C^{E_6}_3 + \frac{1}{9} C^{E_6}_1 + C^{\prime}_1 + \Delta_{T_{1,2}}+ \Delta^\prime_{T_{1,2}} + \frac{2}{5}K + \frac{1}{10}K^{\prime} + 2 g_6^2 D .\label{eq:T}
\end{equation}
We now have sufficient equations to solve for the six unknowns $m_{\mathbf{27}}$, $D_N$, $M_{1/2}$, $\cos 2\beta$, $K$ and $g_6^2 D$, which, as in the previous cases, are now fully determined by the low energy sfermion masses. 
\begin{equation}
\begin{pmatrix}
  M^2_{\tilde{u}_L}\\
  M^2_{\tilde{d}_L}\\
  M^2_{\tilde{e}_R}\\
  M^2_{\tilde{u}_R}\\
  M^2_{\tilde{d}_R}\\
  M^2_{\tilde{T}_{1,2}}
\end{pmatrix} =
\begin{pmatrix}
1 &  c_{\tilde{u}_L} & \delta_{\tilde{u}_L} & -\frac{1}{5} & -1 & - 1\\
1 &  c_{\tilde{d}_L} & \delta_{\tilde{d}_L} & -\frac{1}{5} & -1 & - 1\\
1 &  c_{\tilde{e}_R} & \delta_{\tilde{e}_R} & -\frac{6}{5} & -1 & - 1\\
1 &  c_{\tilde{u}_R} & \delta_{\tilde{u}_R} &  \frac{4}{5} & -1 & - 1\\
1 &  c_{\tilde{d}_R} & \delta_{\tilde{d}_R} & -\frac{2}{5} & -2 & 2  \\
1 &  c_{\tilde{T}_{1,2}} & \delta_{\tilde{T}_{1,2}} & \frac{2}{5} & 2 & 2
\end{pmatrix}
\begin{pmatrix}
  m^2_{\mathbf{27}}\\
  M^2_{1/2}\\
  \cos 2\beta\\
  K\\
  D_N\\
  g_6^2 D
\end{pmatrix}.
\label{eq:E6}
\end{equation}
This provides us with the same results as before for $M_{1/2}$, $\cos 2\beta$ and $K$, and the additional expressions
\begin{eqnarray}
m_{27}^2 &=& \frac{1}{3 X_5} \left[ -c_{\tilde{u}_R}(M_{\tilde{d}_L}^2 + M_{\tilde{T}_{1,2}}^2 + M_{\tilde{u}_L}^2) - c_{\tilde{e}_R}(M_{\tilde{d}_L}^2 + M_{\tilde{T}_{1,2}}^2 + M_{\tilde{u}_L}^2)\right.\nonumber\\ 
&& \left. + (c_{\tilde{d}_L} + c_{\tilde{u}_L})(M_{\tilde{T}_{1,2}}^2 + M_{\tilde{e}_R}^2 + M_{\tilde{u}_R}^2) + c_{\tilde{T}_{1,2}}(-M_{\tilde{d}_L}^2 + M_{\tilde{e}_R}^2 - M_{\tilde{u}_L}^2 + M_{\tilde{u}_R}^2) \right], \label{eq:m27}\\
D_N &=&  \frac{1}{20 X_5} \left[ c_{\tilde{e}_R}(-2 M_{\tilde{d}_L}^2 + 5 M_{\tilde{d}_R}^2 - 5 M_{\tilde{T}_{1,2}}^2 - 2  M_{\tilde{u}_L}^2 + 4  M_{\tilde{u}_R}^2 ) \right.\nonumber \\
&& \left. + c_{\tilde{u}_R}(2 M_{\tilde{d}_L}^2 + 5 M_{\tilde{d}_R}^2 - 5 M_{\tilde{T}_{1,2}}^2 - 4  M_{\tilde{e}_R}^2 + 2  M_{\tilde{u}_L}^2 )\right.\nonumber\\ 
&& \left. + (c_{\tilde{d}_L} + c_{\tilde{u}_L})( -5 M_{\tilde{d}_R}^2 + 5 M_{\tilde{T}_{1,2}}^2 + 2 M_{\tilde{e}_R}^2 -2 M_{\tilde{u}_R}^2)\right.\nonumber\\ 
&& \left. +( c_{\tilde{d}_R} - c_{\tilde{T}_{1,2}})(M_{\tilde{d}_L}^2 - M_{\tilde{e}_R}^2 + M_{\tilde{u}_L}^2 - M_{\tilde{u}_R}^2) \right]  .\label{eq:DN}\\
g_6^2 D &=&  \frac{1}{12 X_5} \left[ c_{\tilde{e}_R}(2 M_{\tilde{d}_L}^2 - 3 M_{\tilde{d}_R}^2 - M_{\tilde{T}_{1,2}}^2 + 2  M_{\tilde{u}_L}^2) + c_{\tilde{u}_R}(2 M_{\tilde{d}_L}^2 - 3 M_{\tilde{d}_R}^2 - M_{\tilde{T}_{1,2}}^2 + 2  M_{\tilde{u}_L}^2 )\right.\nonumber\\ 
&& \left. + (c_{\tilde{d}_L} + c_{\tilde{u}_L})( 3 M_{\tilde{d}_R}^2 + M_{\tilde{T}_{1,2}}^2 - 2 M_{\tilde{e}_R}^2 - 2 M_{\tilde{u}_R}^2) \right.\nonumber\\ 
&& \left.+( 3 c_{\tilde{d}_R} + c_{\tilde{T}_{1,2}})( - M_{\tilde{d}_L}^2 + M_{\tilde{e}_R}^2 - M_{\tilde{u}_L}^2 + M_{\tilde{u}_R}^2) \right]  .\label{eq:g6D}
\end{eqnarray}

The sum rule of equation (\ref{eq:bisr1}) remains unchanged since the extra $E_6$ contributions cancel (in particular $\tilde{u}_L$ and $\tilde{e}_L$ have the same $U(1)_N$ charges as $\tilde{d}_L$ and $\tilde{\nu}_L$ respectively). However, equations (\ref{eq:Sum11}-\ref{eq:Sum12}) are changed by the presence of extra matter. Eliminating $m^2_{\varphi}(0)$, $\Delta_{\varphi}$, $K$ and $D_N$ from equations (\ref{eq:essm1}-\ref{eq:essm7}), we find
\begin{equation}
m^2_{\tilde{u}_L} + m^2_{\tilde{d}_L} - m^2_{\tilde{u}_R} - m^2_{\tilde{e}_R}= C^{E_6}_3 + 2C^{E_6}_2 - \frac{25}{18}C^{E_6}_1 - \frac{3}{4}C^{\prime}_1 \approx 2.8 M^2_{1/2},\label{eq:Sum21e6} 
\end{equation}
and
\begin{equation}
\frac{1}{2}\left(m^2_{\tilde{u}_L} + m^2_{\tilde{d}_L}-m^2_{\tilde{e}_L} -m^2_{\tilde{\nu}_L}\right)  +m^2_{\tilde{d}_R} - m^2_{\tilde{e}_R} 
= 2C^{E_6}_3 - \frac{10}{9}C^{E_6}_1 - \frac{3}{4}C^{\prime}_1 \approx 4.4 M^2_{1/2}.\label{eq:Sum22e6}
\end{equation}
these are considerably different from the sum rules for $SU(5)$, $SO(10)$ and $E_6$ (with no extra matter) so should allow us to distinguish the $E_6SSM$ even without seeing the additional exotic $T_{1,2}$, $\bar T_{1,2}$ or their scalar partners.

\section{A Comparison with SOFTSUSY} \label{sec:softsusy}

In this Section, we will check that the $SU(5)$ and $SO(10)$ sum rules obtained from the one-loop RGEs for the first and second generations, are consistent with the results arising from SOFTSUSY 3.3.0 \cite{Allanach:2001kg}, when $SU(5)$ and $SO(10)$ boundary conditions are imposed. This will then assess the impact of including the full Yukawa couplings as well as the two-loop corrections. We will not compare the $E_6SSM$ sum rule results, since this requires the implementation of new RGEs into SOFTSUSY. While this is in principle available  (see Ref.~\cite{Athron:2009bs}) we leave this for a future study.

\subsection{$SU(5)$ Boundary Conditions}

To test the sum rules of Eqs.~(\ref{eq:Sum11}) and (\ref{eq:Sum12}), one would like to fix all but one of the sparticle masses on the left-hand-side of the equations. One could then vary $M_{1/2}$ and compare the remaining mass prediction from the sum rule with the equivalent prediction from SOFTSUSY including two loop running and a full dependence on the Yukawa couplings. This would tell us how robust these sum rules are under removal of the assumptions used to provide an analytic solution. However, since all of the masses on the left-hand-sides are outputs of SOFTSUSY, this is rather tricky to do. Instead, we define, 
\begin{eqnarray}
\Sigma_1 &\equiv&
M^2_{\tilde{u}_L} + M^2_{\tilde{d}_L}  - M^2_{\tilde{u}_R} - M^2_{\tilde{e}_R}, \\
\Sigma_2 &\equiv& 
\frac{1}{2}\left(M^2_{\tilde{u}_L} + M^2_{\tilde{d}_L} - M^2_{\tilde{e}_L} - M^2_{\tilde{\nu}_L}\right)
+ M^2_{\tilde{d}_R} - M^2_{\tilde{e}_R} ,
\end{eqnarray}
so that the sum rules become,
\begin{eqnarray}
\Sigma_1 &=&  4.8 M^2_{1/2}, \label{eq:y1}\\
\Sigma_2 &=& 8.1 M^2_{1/2}. \label{eq:y2}
\end{eqnarray}
Now we fix all the input parameters except for $M_{1/2}$ and compare the predictions for $\Sigma_1$ and $\Sigma_2$ both from these simple sum rules and from SOFTSUSY as $M_{1/2}$ is varied.

The required inputs are $\tan \beta$ and the boundary conditions at the unification scale. For $SU(5)$, these are the common scalar masses $m_{\overline{\mathbf{5}}}$, $m_{\mathbf{10}}$, $m_{\overline{\mathbf{5}}^{\prime}}$ and $m_{\mathbf{5}^{\prime}}$, the common universal gaugino mass $M_{1/2}$, and a common universal trilinear coupling $A_0$. Note that the choice of $A_0$ is unimportant, since the contributions from trilinear terms are negligible for the first and second generations. However, one should ensure that the choice of $A_0$ does not generate an unstable vacuum; a safe choice is to set $A_0 = 0$. For $m_{\overline{\mathbf{5}}}$ and $m_{\mathbf{10}}$ and $\tan \beta$, we choose $SU(5)$ inputs that generate the masses of scenarios 1, 2 and 3 we already examined in Section~\ref{sec:boundary_conditions}. These $SU(5)$ inputs are shown in Table \ref{table:mass}. The Higgs masses $m_{\overline{\mathbf{5}}^{\prime}}$ and $m_{\mathbf{5}^{\prime}}$ are related through the parameter $S_0$, and we choose to fix $S_0$ to reproduce the three scenarios. Then, the only additional input required is $m_{\overline{\mathbf{5}}^{\prime}}$, which wasn't needed in the earlier analysis. The chosen values for $m_{\overline{\mathbf{5}}^{\prime}}$ are given in Table~\ref{table:mass} and then $m_{\mathbf{5}^{\prime}}$ is fixed by,  
\begin{equation}
 m_{\mathbf{5}^{\prime}} = \sqrt{S_0 + m^{2}_{\overline{\mathbf{5}}^{\prime}}}. \label{eq:m5prime}
\end{equation}
We make no attempt to constrain $m_{\overline{\mathbf{5}}^{\prime}}$ here using the LHC Higgs mass constraints~\cite{:2012gk} since our only motivation is to show that these sum rules are robust to the inclusion of higher orders and the Yukawa couplings. 

\begin{figure}[ht]
\vspace{8mm}
        \begin{subfigure}[b]{0.5\textwidth}
		~~~~~Scenario 1\\[-6mm]
                \centering
                \includegraphics[width=\textwidth]{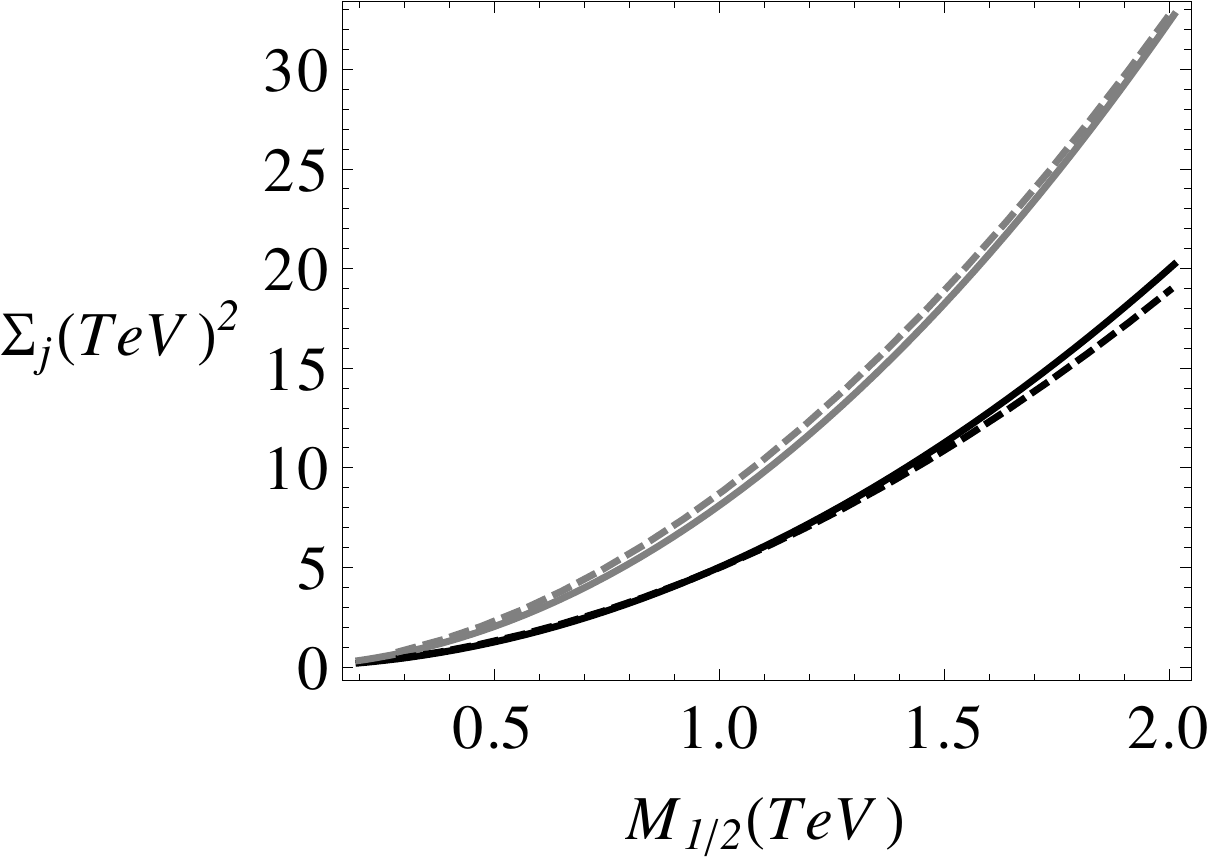}
                \label{fig:su5a}
        \end{subfigure}%
        ~ 
        \begin{subfigure}[b]{0.5\textwidth}
		~~~~~Scenario 2\\[-6mm]
                \centering
                \includegraphics[width=\textwidth]{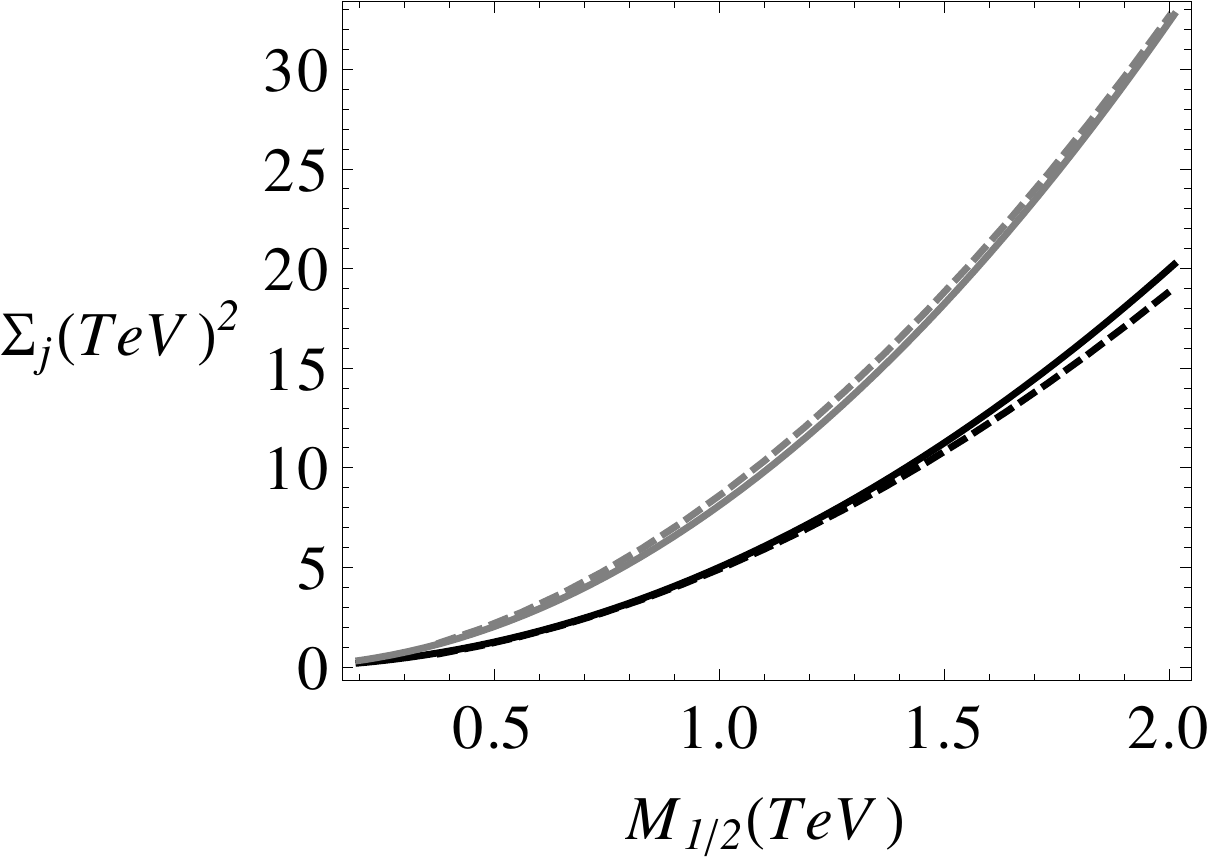}
                \label{fig:su5b}
        \end{subfigure}
        
\begin{center}
        \begin{subfigure}[b]{0.5\textwidth}
		~~~~~Scenario 3\\[-6mm]
                \centering
                \includegraphics[width=\textwidth]{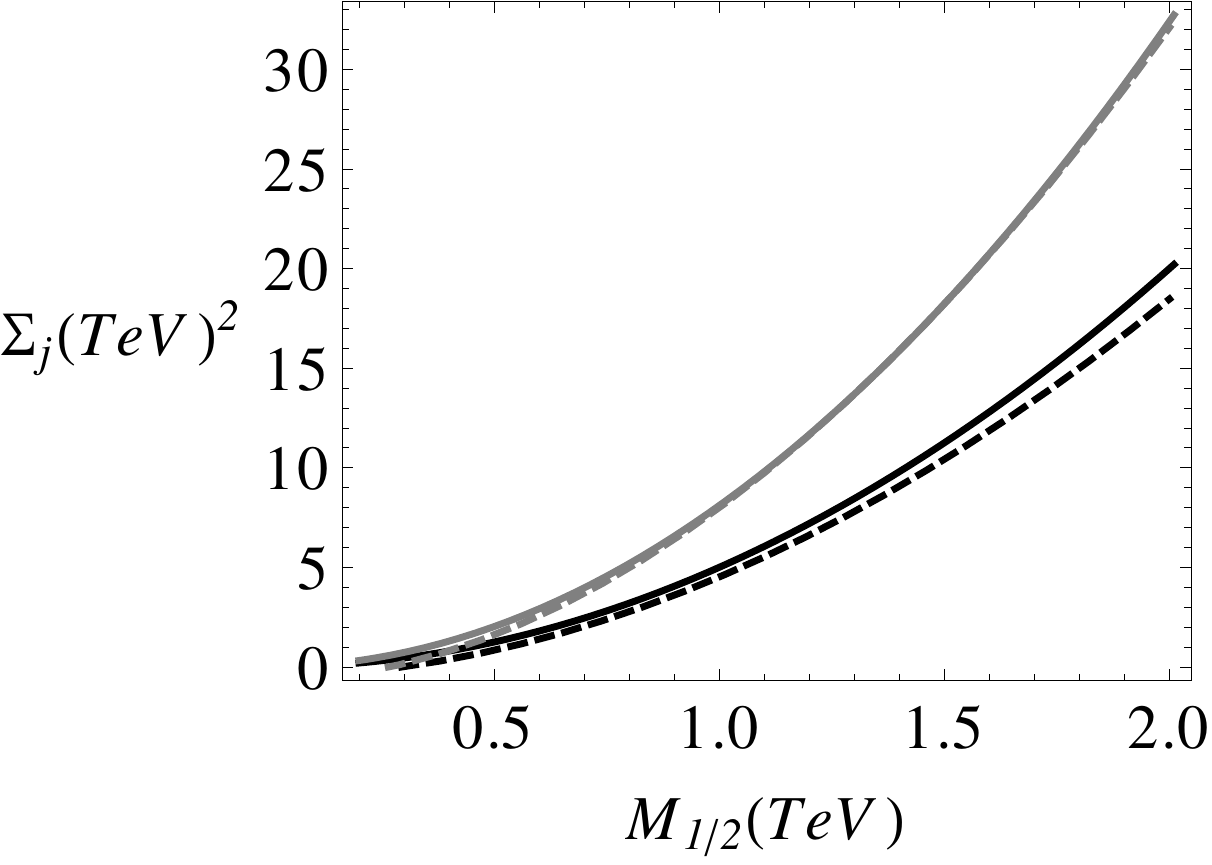}
                \label{fig:su5d}
        \end{subfigure}
        \caption{A comparison of the $SU(5)$ analytic sum rules with SOFTSUSY for example scenarios 1, 2 and 3. The lower solid line is the sum rule of equation (\ref{eq:y1}) while the upper solid line is that for equation (\ref{eq:y2}). The corresponding dashed lines are the results obtained from SOFTSUSY.}\label{fig:su5}
\end{center}
\end{figure}

The results are shown in Figure~\ref{fig:su5}, where the solid lines are the sum rules of equations (\ref{eq:y1} - \ref{eq:y2}) and the corresponding dashed lines are the results obtained from SOFTSUSY. We observe good agreement between the analytic sum rules and the masses obtained from SOFTSUSY at two-loops, indicating that these sum rules are robust.

\subsection{$SO(10)$ Boundary Conditions}

\begin{figure}[ht]
        \begin{subfigure}[b]{0.5\textwidth}
		~~~~~Scenario 1\\[-6mm]
                \centering
                \includegraphics[width=\textwidth]{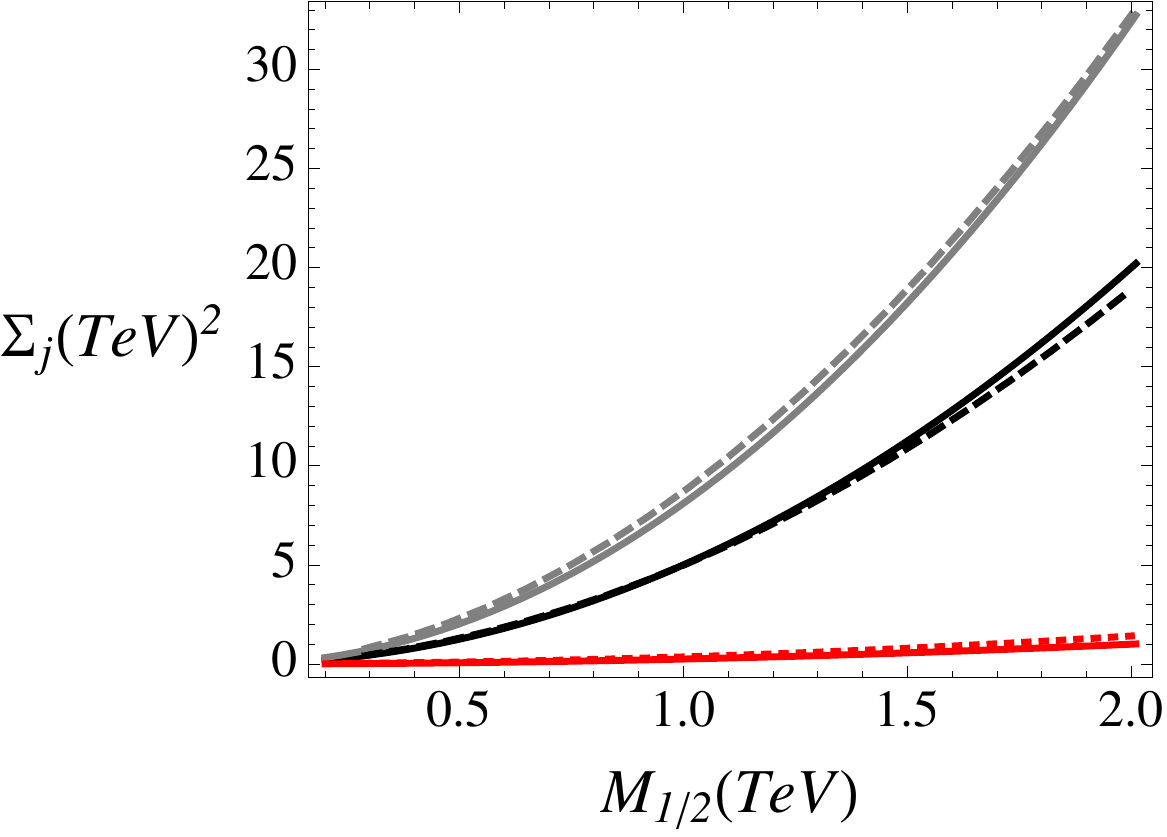}
                \label{fig:so10a}
        \end{subfigure}%
        ~ 
        \begin{subfigure}[b]{0.5\textwidth}
		~~~~~Scenario 2\\[-6mm]
                \centering
                \includegraphics[width=\textwidth]{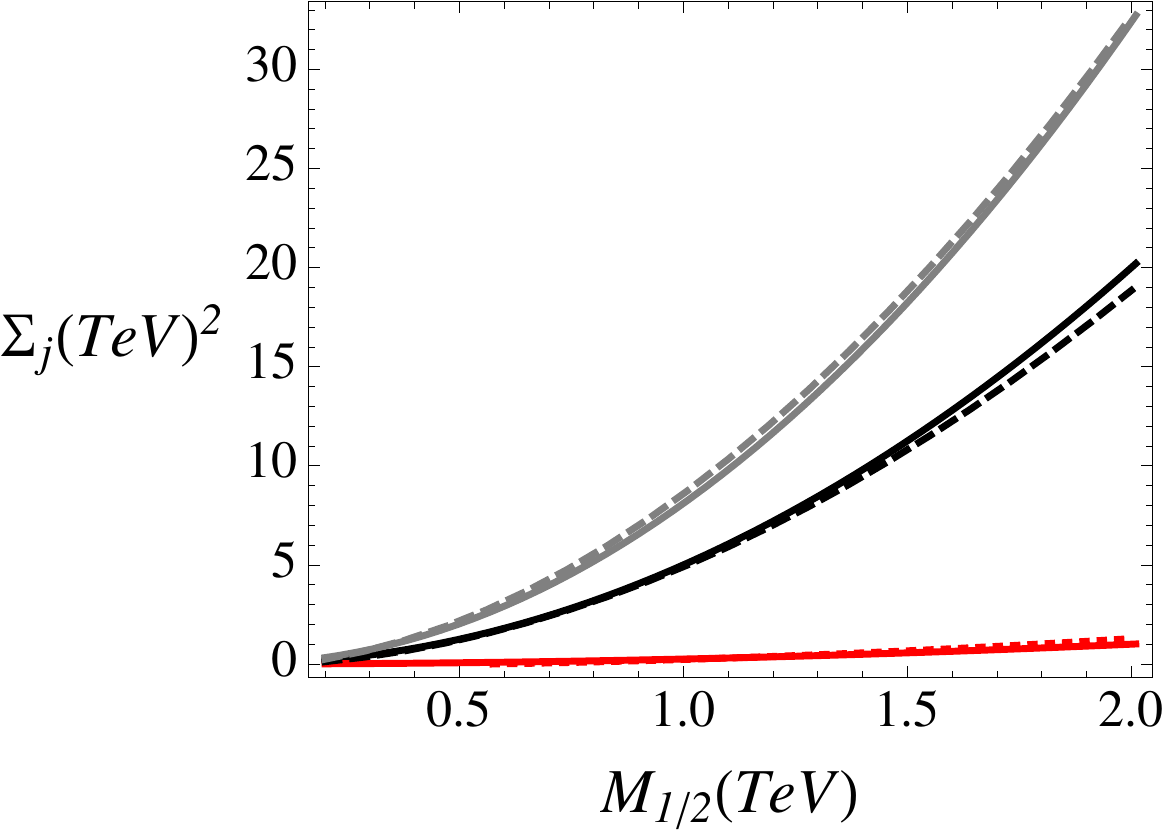}
                \label{fig:so10b}
        \end{subfigure}
        
\begin{center}
        \begin{subfigure}[b]{0.5\textwidth}
		~~~~~Scenario 3\\[-6mm]
                \centering
                \includegraphics[width=\textwidth]{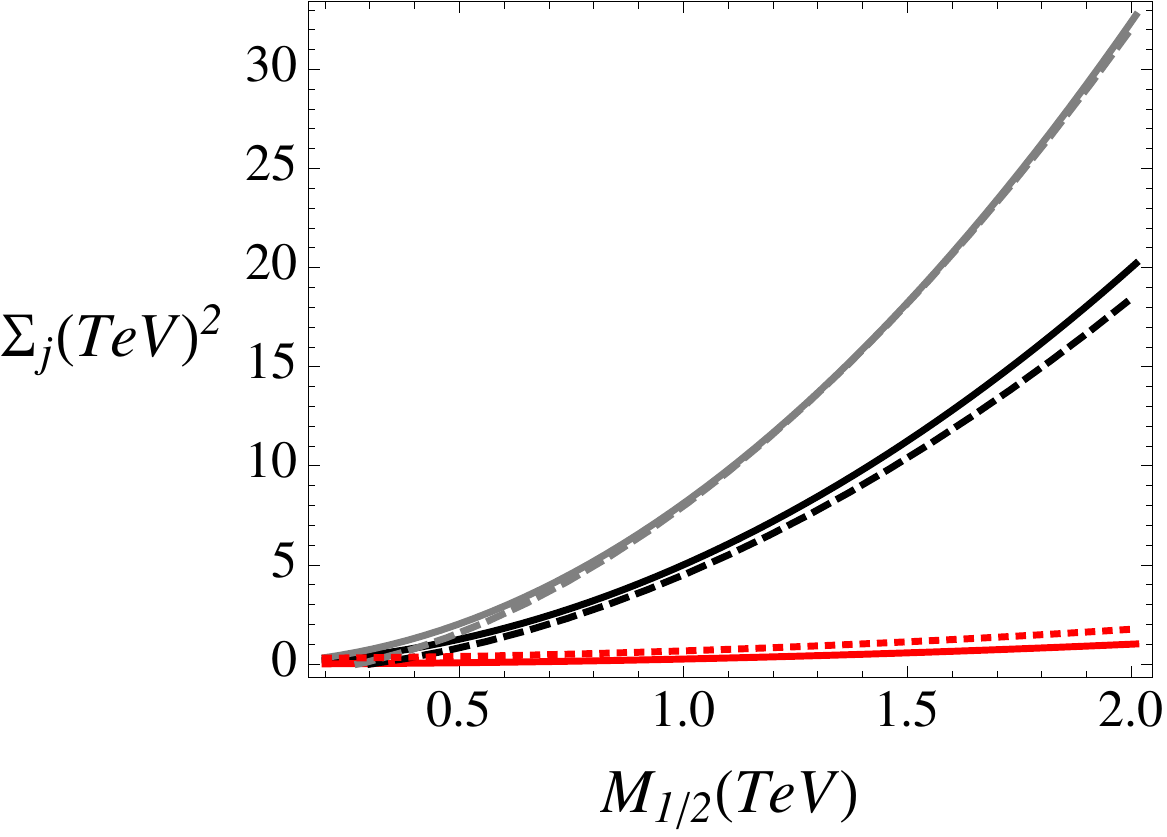}
                \label{fig:so10c}
        \end{subfigure}
        \caption{A comparison of the $SO(10)$ analytic sum rules with SOFTSUSY for example scenarios 1, 2 and 3. The lower solid line is now the sum rule of equation (\ref{eq:y3}), while the middle solid line and the upper solid line are the sum rules given by equations (\ref{eq:y1}) and (\ref{eq:y2}) respectively. The corresponding dashed lines are the results obtained from SOFTSUSY.}\label{fig:so10}
\end{center}
\end{figure}

We also test the sum rules for $SO(10)$ boundary conditions. Now, in addition to $\tan \beta$, we have a common mass for the sfermions, $m_{\mathbf{16}}$, a common mass for the Higgs, $m_{\mathbf{10}}$, and D-term arising from the breaking of $SO(10)$, $g^2_{10} D$. As before, we chose our inputs, $m_{\mathbf{16}}$, $\tan \beta$ and $g^2_{10} D$ such that they reproduce our example scenarios. Again, the common Higgs mass wasn't needed for the earlier examples, but now we must fix it within SOFTSUSY and use the values given in Table (\ref{table:mass}). The results of this analysis are the two upper sets of curves in Figure (\ref{fig:so10}). Once again, the analytic sum rules are in good agreement with SOFTSUSY.

We saw earlier that $SO(10)$ also implied an extra constraint, equation (\ref{eq:constraint}), which relates $K$ (and therefore $S_0$) to the D-term. Since $K$ and $g^2_{10}D$ are both functions of the low energy masses, equations (\ref{eq:K}) and (\ref{eq:D10}), this provides us with an additional sum rule. As for the previous sum rules, this form is a little hard to check in SOFTSUSY since both sides of the equation are outputs of SOFTSUSY. We therefore first make a few manipulations to bring an input on to one side of the equation, allowing us to vary the input and check the robustness of the sum rule. Substituting equation (\ref{eq:m12}) into (\ref{eq:D10}) one can write, 
\begin{equation}
g_{10}^2D =  \mathfrak{D} + 5c_{\tilde{d}_R} \frac{M^{2}_{1/2}}{20},
\end{equation}
where,
\begin{eqnarray}
 \mathfrak{D}  &\equiv& \frac{1}{20X_5} \left[ -c_{\tilde{u}_R}(2M_{\tilde{d}_L}^2 - 5M_{\tilde{d}_R}^2 + M_{\tilde{e}_R}^2 + 2M_{\tilde{u}_L}^2) - c_{\tilde{e}_R}(-3M_{\tilde{d}_L}^2 + 5M_{\tilde{d}_R}^2 - 3M_{\tilde{u}_L}^2 + M_{\tilde{u}_R}^2)\right. \nonumber\\ 
&& \qquad\quad \left. + (c_{\tilde{d}_L} + c_{\tilde{u}_L})(5M_{\tilde{d}_R}^2 - 3M_{\tilde{e}_R}^2 - 2M_{\tilde{u}_R}^2) \right] \label{eq:curlyD}.
\end{eqnarray}
Substituting this back into the constraint, equation (\ref{eq:constraint}), and rearranging to place $M_{1/2}$ on the right-hand side, we find,
\begin{eqnarray}
\Sigma_3 &=& \frac{1}{4} M^2_{1/2}, \label{eq:y3}
\end{eqnarray}
where,
\begin{equation}
\Sigma_3 \equiv \frac{1}{c_{\tilde{d}_R} } \left( -\frac{1}{2}b_1K\left[ 1 - \frac{\alpha_1(t)}{\alpha_1(0)} \right]^{-1}
- \mathfrak{D} \right).
\end{equation}
All the masses in $K$ and $\mathfrak{D}$, and hence $\Sigma_3$, are outputs, so the sum rule may be compared with SOFTSUSY as the input $M_{1/2}$ is varied. These comparisons are shown as the lower set of curves in Figure~\ref{fig:so10}, where the solid curve is the simple analytic expression and the dashed curve is the SOFTSUSY result. Once again we have good agreement indicating that these rules are robust.

\section{The 125 GeV Higgs candidate and fine-tuning} \label{sec:higgs}

The recent observation of a 125 GeV Higgs candidate at the LHC~\cite{:2012gk}, raises further questions. In general, the Higgs boson has little effect on the  first and second generation spectrum since the corresponding Yukawa couplings are very small. However, it does constrain the third generation, so in models with a unified high scale physics, it will naively restrict the GUT scale scalar mass from which the first and second generation masses must be run. A 125 GeV Higgs is approaching the limit of feasibility for cMSSM models; to provide such a heavy Higgs one needs a large stop mass, so in a $SU(5)$ GUT one might expect to need a value of  $m_{10}$ approaching a few TeV. However, this is not the full story, since other parameters, such as the gaugino mass also provide a significant contribution to the stop mass, and may be enough to provide a heavy enough stop to force a 125 GeV Higgs without requiring a large value of $m_{10}$. The interplay of the various parameters quickly becomes rather complicated and is beyond the scope of this paper, so we reserve this topic for a future publication. 

It seems fairly generally true that, irrespective of the source of the stop mass, the thus far absence of supersymmetry at the LHC indicates that the supersymmetric spectrum, if it exists, must be heavier than was once hoped. This leads to the models requiring some degree of fine tuning in order to get the correct Z-boson mass, doing damage to one of supersymmetry's most compelling motivations. We close with two further comments on this matter. Firstly, let us not throw the baby out with the bath water; even a multi-TeV supersymmetric spectrum is less fine tuned than the Standard Model and its hierarchy problem. Supersymmetric models, and in particular supersymmetric GUTs have many desirable features and solve so many problems that we should continue to explore their possibility. Secondly, fine-tuning should not be used as a razor to remove misbehaving theories, but as an indicator of where theories are incomplete and can be improved. With this in mind, an observation of a multi-TeV supersymmetry spectrum, most probably with the first and second generation sfermions, would provide an opportunity to probe the high scale physics using the techniques described here, possibly leading to new theorectical developments and a physical explanation of why the Z-boson mass and supersymmetry in general is not fine-tuned after all. 

\section{Discussion and Conclusion} \label{sec:conc}

\hspace{5 mm} In this paper we have studied the RGEs of the sfermion masses of the first and second generations for $SU(5)$, $SO(10)$ and $E_6$ boundary conditions. Neglecting Yukawa couplings in the one-loop RGEs for the first two generations allows an analytical analysis. The parameters of the underlying theory were determined as explicit functions of the low scale squark and slepton masses. An $SO(10)$ supersymmetric GUT, with the choice of Higgs fields in a $\mathbf{10}$ dimensional representation, provides a further constraints on the low scale masses when compared to $SU(5)$. A simplistic $E_6$ model that breaks to $SO(10) \otimes U(1)$ at the GUT scale, with no extra matter below the GUT scale, presents a similar picture to $SO(10)$ only with $m_\mathbf{16}^2$ replaced with the combination $m_\mathbf{27}^2+2g_6^2D_S$. The same analysis was also done for the $E_6SSM$, where an extra $U(1)$ and additional matter survives down to the electroweak scale.  These new effects alter the RGEs as well as introduce new D-terms, at both the GUT and electroweak scales.

The possibility of preforming an analytical study of the RGEs of the first and second families allowed us to obtain sum rules for the different models, and we observe that the $E_6SSM$ is clearly distinguishable from the other three cases. These sum rules can therefore be used to quickly identify the GUT gauge group from the spectrum of the first two generations.

Of course, the underlying GUT scale parameters will also affect the RGEs of the third generation. Analytic expressions for these parameters in terms of the low scale first or second generation masses allow one to use the first or second generation masses as inputs to the analysis of the third generation. Since the experimental constraints on the supersymmetric parameter space come mainly from the first and second generation, using these as inputs may provide a more efficient methodology for exploring the third generation parameter space of grand unified models.

\section*{Acknowledgements}

P.N.P.~would like to thank the University of Glasgow and D.~J.~Miller for hospitality during a visit to Glasgow where this work was started. The work of P.N.P. is supported by the J.~C.~Bose National Fellowship and by the Council of Scientific and Industrial Research, India under project No.~03(1220)/12/EMR-II. D.J.M.~acknowledges partial support from the STFC Consolidated Grant
ST/G00059X/1. A.M.~would like to acknowledge FCT for the grant SFRH/BD/62203/2009 under the strategic project PEst-C/FIS/LA0007/2011. A.M.~and D.J.M.~would also like to thank Dr.~David Sutherland for his constant support, criticism and fruitful discussions in the realisation of the work presented in this paper.

\end{document}